\documentclass[intlimits,twoside,a4paper]{article}

\usepackage{amsmath,amssymb}
\usepackage{graphicx}
\usepackage[T2A]{fontenc}
\usepackage[cp1251]{inputenc}
\usepackage[tight]{subfigure} 

\usepackage[eqsecnum]{cmpj3}

\issue{2019}{22}{1}{13701}
\doinumber{10.5488/CMP.22.13701}

\title[Emergence of topological Mott insulators in proximity of quadratic band touching points]%
{Emergence of topological Mott insulators in proximity of quadratic band touching points}
\author[I. Mandal, S. Gemsheim]{I. Mandal\refaddr{label1}, S. Gemsheim\refaddr{label2}}
\addresses{
\addr{label1} Laboratory of Atomic And Solid State Physics, Cornell University, Ithaca, NY 14853, USA 
\addr{label2} Fakult\"at Physik, Technische Universit\"at Dresden, 01069 Dresden, Germany
}

\newcommand{\nn}{\nonumber \\}
\newcommand{\be}{\begin{equation}}
\newcommand{\ee}{\end{equation}}

\renewcommand{\vec}[1]{{\bf #1}}

\renewcommand{\epsilon}{\varepsilon}

\def\be{\begin{eqnarray}}
\def\ee{\end{eqnarray}}

\date{Received October 16, 2018, in final form December 1, 2018}

\begin{document}

\maketitle

\begin{abstract}
Recently, the field of strongly correlated electrons has begun an intense search for a correlation induced topological insulating phase. An example is the quadratic band touching point which arises in a checkerboard lattice at half-filling, and in  the presence of interactions gives rise to topological Mott insulators. In this work, we perform a mean-field theory computation to show that such a system shows instability to topological insulating phases even away from half-filling (chemical potential $\mu  =  0 $). The interaction parameters consist of on-site repulsion ($ U $), nearest-neighbour  repulsion ($ V $), and a next-nearest-neighbour correlated hopping ($ t_\text{c} $). The $t_\text{c}$ interaction  originates  from  strong  Coulomb  repulsion. By tuning the  values of these parameters, we obtain a desired topological phase that spans the area around $(V  =  0  ,  \mu  =  0)$, extending to regions with $(V>0,\mu=0)$ and $(V>0,\mu>0)$. This extends the realm of current experimental efforts to find these topological phases.
\pacs 73.43.-f, 73.43.Nq, 71.10.Fd
\end{abstract}

\section{Introduction}
Study of topological phases in condensed matter systems is one of the most active areas of research in recent times~\cite{hasan-kane}.
In conventional topological insulators, a combination of spin-orbit interactions and time-reversal symmetry gives rise to protected conducting states at edges/surfaces in spite of the presence of a bulk band gap (like an ordinary insulator). The focus has mostly been on noninteracting systems. In this work, we consider the proposal of inducing topological insulating phases in 2D materials through interactions, without the need for spin-orbit coupling or large intersite interactions~\cite{raghu,kai-sun,herbut,tsai,vafek,wu,stephan-review}. While bulk insulating gaps arise due to interactions, the topological nature is captured by the topologically protected edge states. We will dub them ``topological Mott insulators''~\cite{raghu}. The quantum anomalous Hall (QAH) effect, emerging as a 2D topological insulating phase, can be understood as a generalization of the quantum Hall effect for the spin-singlet case, which is an integer quantum Hall phase with gapless chiral currents at the edges, but realized in the absence of any external magnetic field. The QAH ground state breaks time reversal symmetry with unbroken lattice translational symmetry and has a bulk insulating gap.
The quantum spin Hall (QSH) effect is the analogue of the QAH effect, but the gapless edge states are helical such that electrons with the opposite spins counter-propagate giving rise to spin currents (rather than charge currents). Moreover, the ground state does not break time reversal symmetry.
In the set-ups for realizing interaction-driven 2D topological phases proposed so far~\cite{kai-sun,herbut,tsai,vafek,wu}, electron-electron interactions at a quadratic band crossing point (QBCP) were considered, because parabolically touching bands have a finite density of states in 2D.

QBCPs can arise on the checkerboard~\cite{kai-sun} (at $\frac{1}{2}$ filling), Kagome~\cite{kai-sun} (at $\frac{1}{3}$ filling),  and  Lieb~\cite{tsai} lattices. The spin-singlet $d$-density wave (DDW) state in a checkerboard lattice is the same as the QAH phase discussed in \cite{kai-sun,dona}. This is because when the diagonal hopping terms are modulated right at the lattice level, this in effect gives us the $d_{xy} + \ri d_{x^2-y^2}$ phase on the square lattice, which is the QAH phase involving the time-reversal symmetry breaking current loops. The corresponding triplet variety is the QSH phase. We also note that the QSH phase has the same energy as the QAH phase, and hence cannot be distinguished at the mean-field level.

The importance behind the study of checkerboard lattice is as follows: This lattice
consists of criss-crossed squares, which can be viewed as a 2D projection of a 3D pyrochlore lattice (whose structure is very common in nature) onto a plane. Furthermore, in the cuprate superconducting materials, each copper-oxygen layer forms a checkerboard lattice consisting of alternate copper and oxygen ions, such that the oxygen ions form squares with copper ions at the centers. There is one orbital per site, resulting in two bands crossing at a QBCP at the wavevector $ (\piup,\piup)$ with a fourfold rotational symmetry. Hence, at half filling, the Fermi level and the QBCP coincide.

We must emphasize that checkerboard lattice is merely a 2D analogue of the 3D pyrochlore lattice. Similarly, the cuprate planes are stacked in 3D to form a complex 3D structure. A change in dimensionality, especially in presence of a Fermi surface (not a Fermi point), can lead to nontrivial consequences. Hence, there is a caveat in naively assuming checkerboard lattices to explain these real materials. Nonetheless, as with other simplifying physics models, we can consider the checkerboard lattice as a toy model to get some real intuition about the actual physics going on in the 3D systems. The real motivation of studying it in this work is of course to analyse the phases for QBCP and at points away from the exact QBCP. 

There is no intrinsic reason why the interaction-induced QAH phase should not exist for a generic $\mu$ if the lattice and the interactions are of the right kind.
In the previously studied models, the interaction parameters considered were: on-site repulsion ($ U $), and nearest-neighbour  repulsion ($ V $). We generalize this by including a pair-hopping (or correlated hopping)~\cite{Nayak:2002}, denoted by an interaction strength $t_\text{c}$, as this term is known to favour DDW/QAH ordering~\cite{Nayak:2000}. The next-nearest-neighbour correlated hopping ($ t_\text{c} $) originates from strong  Coulomb  repulsion. In this paper, our goal is to show that there is a QAH phase on the checkerboard lattice (at least in mean-field theory) for a chemical potential ($\mu$) that does not exactly correspond to a Fermi point at quadratic band touching, and in fact it continuously connects/extends to the regions of $(V=0,\mu=0)$ and $(V>0, \mu=0)$, given that we tune $U$ and $t_\text{c}$ to some optimal values.
Our study considers instabilities among QAH, charge-density wave (CDW) and spin-density wave (SDW) phases.
We employ a mean-field theory approach by minimizing the free energy involving the possible competing phases in order to study the effect of chemical potentials away from the QBCP.

%%%%%%%%%%%%%%%%%%%%%%%%%%%%%%%%%%%%%%%%%%%%

\section{The checkerboard lattice model}
\label{model}

\begin{figure}[!b]
\centering
\includegraphics[width =0.37  \linewidth]{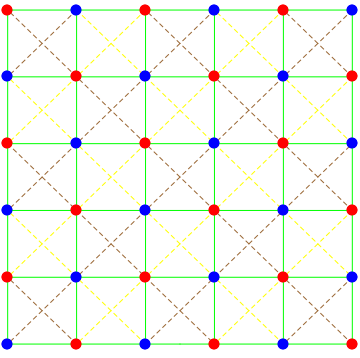}
\caption{\label{lattice} (Colour online) The checkerboard lattice, with the nearest-neighbour hopping of amplitude~$t$ denoted by the green lines. The two next-nearest-neighbour hopping amplitudes $\frac{t_\text{diag}}{2}$ and $-\frac{t_\text{diag}}{2}$ are shown by dashed brown and dashed yellow lines, respectively.}
\end{figure}

We consider the Hamiltonian of \cite{wu}, illustrated in figure~\ref{lattice}:
\begin{align}
H &= H_{0}  + H_\text{int}\,,\nonumber
\end{align}
\begin{align}
H_{0}  & = - \mu  \sum \limits_{\sigma} \sum \limits_{i,j }
 c_{i,j; \sigma}^\dagger c_{ i,j;\sigma}
-   t  \sum \limits_{\sigma} \sum \limits_{ i,j } 
\left (  c_{ i,j; \sigma}^\dagger c_{ i+1,j;\sigma} +  c_{ i,j; \sigma}^\dagger c_{ i,j+1;\sigma} 
+ c_{ i,j; \sigma}^\dagger c_{ i-1,j;\sigma} +  c_{ i,j; \sigma}^\dagger c_{ i,j-1;\sigma} 
\right ) \nn
& \quad  + \frac{t_\text{diag}}{2}  \sum \limits_{\sigma} \sum \limits_{i,j } (-1)^{i+j} \left (
c_{i,j; \sigma}^\dagger c_{ i+1,j+1;\sigma} + c_{i,j; \sigma}^\dagger c_{ i-1,j-1;\sigma}
- c_{i,j; \sigma}^\dagger c_{ i+1,j-1;\sigma} - c_{i,j; \sigma}^\dagger c_{ i-1,j+1;\sigma}
\right ),\nn
%%%%%%%%%%%%%%%%%%%%%%%%%%%%%%%%%%%%%%%%%%%%%%%%%%
H_\text{int}  & = U  \sum \limits_{ i,j} n_{i,j;\uparrow} n_{i,j;\downarrow}
+ 
V \sum \limits_\sigma \sum \limits_{ \langle m, m' \rangle} n_{m;\sigma} n_{m';\sigma } 
-t_\text{c}  \sum \limits_{\sigma } \sum \limits_{\substack{ \langle m, m'  \rangle,     \langle m',  m'' \rangle \\ m\neq m''}}
 c_{m; \sigma}^\dagger c_{m';\sigma}  c_{m';\sigma}^\dagger   c_{m'';\sigma}  \,,
 \label{ham1}
\end{align}
with $ \lbrace m=(i,j), m'=(i',j'), m'' =(i'',j'') \rbrace $ and $(\sigma, \sigma')$ denoting the site and spin indices, respectively. The nearest-neighbour pairs hop with strength $t$, while the next-nearest neighbours are connected by diagonal bonds with hopping strength $t_\text{diag}$. We note that the diagonal hoppings give rise to the commensurate singlet $d_{xy}$-density wave ordering \cite{Nayak:2000}.
In the interaction terms, $U $ is the on-site repulsion, $V$ is the nearest-neighbour repulsion, and $t_\text{c}$ is the next-nearest-neighbour correlated hopping. The symbol $ \langle m, m' \rangle$ indicates nearest-neighbour pairs. The next-nearest-neighbour correlated hopping occurs when an electron hops from $m''$ to $m'$ when $m'$, in turn, is vacated by an electron hopping to $m$. For the non-interacting part ($H_0$), the QBCP appears at half-filling ($\mu=0$), when the non-interacting electrons have a finite density of states but lack a Fermi surface. To simplify the notation, $t$ and the nearest-neighbour bond length $a$ are set to be units of energy and length.

Our aim is to reduce the Hamiltonian into a form including various ordered phases~\cite{Nayak:2000, Nayak:2002} as follows:\footnote{Nayak and Pivovarov~\cite{Nayak:2002} solved the mean-field phase diagram arising from correlated hopping on a bilayer square lattice, while we are interested in a single layer checkerboard lattice.}
\begin{align}
H_\text{mf}=&\int{[\rd\mathbf{k}] \left( \epsilon_\mathbf{k} -\mu \right) c^\dagger_{\mathbf{k};\sigma} c_{\mathbf{k};\sigma}} + \int{[\rd\mathbf{k}]\tilde{\epsilon}_\mathbf{k}(c^\dagger_{\mathbf{k}+\mathbf{Q}  ;\sigma}   c_{\mathbf{k};\sigma}+\text{h.c.})}\nonumber \\
&-g_\text{ddw}\iint{[\rd\mathbf{k}][\rd\mathbf{k}']f_\mathbf{k}f_\mathbf{k}'
\left[c^\dagger_{\mathbf{k}+\mathbf{Q};\sigma} c_{\mathbf{k};\sigma} \right]
\left[ c^\dagger_{\mathbf{k}';\sigma'}   c_{\mathbf{k}'+\mathbf{Q};\sigma'}\right] } \nonumber \\
%%%%%%%%%
&-g_\text{ddw}\iint{[\rd\mathbf{k}]   [\rd\mathbf{k}'] f_\mathbf{k} f_\mathbf{k}'\left[c^{\dagger}_{\mathbf{k}+\mathbf{Q};\alpha}\sigma_z^{\alpha\beta}c_{\mathbf{k}; \,\beta} \right]
\left[ c^{\dagger}_{\mathbf{k}';\gamma} \sigma_z^{\gamma\delta} c_{\mathbf{k}'+\mathbf{Q}; \delta}\right ]}\nonumber \\
%%%%%%%%%%%%%%%%%%%%%%%
&-g_\text{sdw}\iint{[\rd\mathbf{k}][\rd\mathbf{k}']\left [c^{\dagger}_{\mathbf{k}+\mathbf{Q};\alpha} \sigma_z^{\alpha\beta}
c_{\mathbf{k} ; \, \beta}\right]
\left[c^{\dagger}_{\mathbf{k}'; \gamma}\sigma_z^{\gamma\delta}c_{\mathbf{k}'+\mathbf{Q}; \delta} \right ]}  \nonumber \\
%%%%%%%%%%%%%%%%%%%%%%%%
& -g_\text{dsc}\iint{[\rd\mathbf{k}][\rd\mathbf{k}']f_\mathbf{k}f_\mathbf{k}'
\left[c^\dagger_{\mathbf{k}\uparrow}c_{-\mathbf{k}\downarrow}\right]
\left[ c^\dagger_{-\mathbf{k}'\uparrow} c_{\mathbf{k}'\downarrow} \right]}
%%%%%%%%%%%%%%%%%%%
\nn &  -g_\text{cdw}\iint{[\rd\mathbf{k}][\rd\mathbf{k}'] 
\left[c^\dagger_{\mathbf{k}+\mathbf{Q};\sigma} c_{\mathbf{k};\sigma} \right]
\left[ c^\dagger_{\mathbf{k}';\sigma'} c_{\mathbf{k}'+\mathbf{Q};\sigma'}\right] }
,\nn
%%%%%%%%%%%%%%%%%%%%%%%%%%%%%%%%%%%%%%%%%%%%%%%%
\mathbf Q = &  (\piup,\piup), \quad f_\mathbf{k} = \cos k_x-\cos k_y\,,\quad [\rd\mathbf{k}] =\frac{\rd k_x \rd k_y}{\piup^2}\,,
\label{ham2}
\end{align}
where the indices $\sigma,\alpha,\beta,\gamma,\delta$ are spin-indices and summation over repeated indices is implied. The free fermionic dispersion on a square lattice is $\epsilon_{\mathbf{k}}=-2 t(\cos k_x + \cos k_y )$, while $\tilde{\epsilon}_\mathbf{k}=-2t_\text{diag} \sin k_x \sin k_y$ is the energy dispersion for the $d_{xy}$ phase emerging from a checkerboard lattice with $t_\text{diag}$ as the diagonal hopping strength on adjacent plaquettes. The third, fourth, fifth, sixth and seventh terms on the RHS represent singlet DDW, triplet DDW, SDW, $d$-wave superconducting (DSC) and CDW phases, respectively. The couplings are given by~\cite{Nayak:2002}:
\begin{align}
& g_\text{ddw}=8V+24 t_\text{c}\,, \quad g_\text{sdw}=2U, \nn
& g_\text{dsc}=12t_\text{c}-8V,\quad g_\text{cdw} = 16V+24t_\text{c} -2U.
\end{align}
Since the order parameters condense at the wave vector $\mathbf Q= (\piup, \piup) $, we use the reduced BZ (RBZ) by folding the full 2D Brillouin zone (BZ). The RBZ is  defined  in  terms  of  the  rotated  coordinates: $k_x' =\frac{k_x+k_y}{\sqrt 2}$, 
$k_y' =\frac{k_x -k_y}{\sqrt 2}$, where $k_x',k_y' \in \big [ -\frac{\piup}{\sqrt 2}, \frac{\piup}{\sqrt 2}\big ]$. 

Here, we will consider only two competing phases --- QAH (or DDW) and SDW. We will assume that the CDW phase is suppressed by a large enough $U$, which is true for the parameter regime when $g_\text{cdw} \leqslant 0$.\footnote{In other words, we will restrict to the regime where CDW is not energetically favourable and therefore ruled out.} We are interested in the regions of strong interactions, where DDW phase can appear, and it is not our motivation to study the regions with all possible phases included. So this criterion considerably shortens the range of parameter space where we have to perform our search, which is computationally expensive. We will also leave out DSC because our aim is to study the phase diagram away from the parameters leading to the familiar high-$T_\text{c}$ cuprate phase diagrams. In other words, we are interested in examining how the topological QAH phase can appear near $\mu=0$ and away from the doping values responsible for DSC. It has already been shown \cite{Nayak:2002} that a DDW phase favours superconductivity in its proximity and will eventually give rise to DSC for large enough $\mu$, and hence this is not what we want to study.

The SDW and the singlet $d_{x^2-y^2}$ order parameters can be expressed as: 
\begin{align}
\label{defn}
&\phi_{\text{sdw}} = g_\text{a}\int [\rd\mathbf{k}]
\left [ \langle c^\dagger_{\mathbf{k}+\mathbf{Q}; \uparrow} c_{\mathbf{k}; \uparrow} \rangle - \langle c^\dagger_{\mathbf{k}+\mathbf{Q}; \downarrow} c_{\mathbf{k}; \downarrow} \rangle    \right ],\nn
&\phi_{\text{ddw}}^s = g_\text{b} \int [\rd\mathbf{k}]f_{\mathbf{k}}\langle 
c^\dagger_{  \mathbf{k}+{ \mathbf{Q} };\sigma  }
c_{\mathbf{k};\sigma}\rangle .
\end{align}
The QAH order parameter is decomposed in momentum space just like a DDW and looks like:
$	\ri g_\text{b} f_{\mathbf{k}}\langle c^\dagger_{\mathbf{k}+\mathbf{Q};\sigma}c_{\mathbf{k};\sigma}\rangle.$

To derive a mean-field theory, it is convenient to take the
Fourier transform of the Hamiltonian in equation~(\ref{ham1}) and regroup the terms. To do so, first let us Fourier transform $H_\text{int}$ to obtain~\cite{Nayak:2000}:
\begin{align}
H_\text{int}  & = \piup^2 \sum \limits_{\sigma,\sigma'}\int   [\rd\vec k_1]     [\rd\vec k_2] [\rd\vec k_3] [\rd\vec k_4] 
\delta ( \vec k_1+ \vec k_3- \vec k_2 -\vec k_4 )  
c^\dagger _{\vec k_1;\sigma}  c _{\vec k_2; \sigma } c^\dagger _{\vec k_ 3;\sigma' }  c _{\vec k_4 ;\sigma'}
\big\{  V  [   \cos ( k^ x _3 - k^ x_4 )  \nn 
&+\cos ( k^ y _3 - k^ y_4 )] 
 -   t_\text{c}[   \cos ( k^ x _1 - k^ x_4 )+\cos ( k^ y _1 - k^ y_4 ) +   2 \cos k ^x _1  \cos k ^x _4 +   2 \cos k ^y _1  \cos k ^y _4 ] \big\} ,
\label{ham-parent}
\end{align}
where $\piup^2 \int   [\rd\vec k]   \delta ( \vec k) =1 .$
As an example, we have demonstrated how to extract the DDW contribution from the  interaction  part  in  appendix~\ref{app}, following \cite{foglio}. In appendix~\ref{order}, we have shown how we can formulate the mean-field theory for various order parameters.

%%%%%%%%%%%%%%%%%%%%%%%%%%%%%%%%%%%

%%%%%%%%%%%%%%%%%%%%%%%%%%%%%%%%%%%%%%%%%%%%%%%%%%%%%%%%%%%%%%%%%%%%%%%%%%%%%%
\section{Phases from free energy minimization}
\label{phase}

At the mean-field level, singlet DDW (or QAH) and triplet DDW (or QSH) turn out to have the same energies and hence we cannot distinguish between them. Therefore, we will determine the phase diagram by considering two order parameters: one for SDW, and the other for singlet DDW.
For notational simplicity, we will use $\phi_\text{a}$ and $\phi_\text{b}$ to denote these two order parameters, respectively. 
The mean-field Hamiltonian, including the above two phases, is given by:
\begin{align}
H_\text{mf}^\text{ab} &=\int [\rd{\mathbf{k}}] \psi^\dagger_{\mathbf{k}}
\left [ h(\mathbf{k})-\mu \right ]
 \psi_{\mathbf{k}}
  + \frac{4|\phi_\text{a}|^2}{g_\text{a}} + \frac{4|\phi_\text{b}|^2}{g_\text{b}} \,,
 \quad \psi^\dagger_{\mathbf{k}}   =\left (c^\dagger_{\mathbf{k}; \uparrow}, c^\dagger_{\mathbf{k}+\mathbf{Q}; \uparrow}, c^\dagger_{\mathbf{k}; \downarrow}, c^\dagger_{\mathbf{k}+\mathbf{Q}; \downarrow}  \right ), \nn
%%%%%%%%%%%%%%%%%%%%%%%%%%%%%%%%%% 
  h(\mathbf{k})&=
  \begin{pmatrix}
  \epsilon_{\mathbf{k}}  &\tilde{\epsilon}_\mathbf{k}-2\phi_\text{a}-2\ri\phi_\text{b}f_\mathbf{k} & 0 & 0\\
  \tilde{\epsilon}_\mathbf{k}-2\phi_\text{a}+2\ri\phi_\text{b}f_\mathbf{k}  & -\epsilon_{\mathbf{k}}  & 0 & 0\\
  0 & 0 &\epsilon_{\mathbf{k}}  & 2 \phi_\text{a}-2\ri\phi_\text{b}f_\mathbf{k} +\tilde{\epsilon}_\mathbf{k}\\
  0 & 0 & 2\phi_\text{a}+2\ri\phi_\text{b}f_\mathbf{k} +\tilde{\epsilon}_\mathbf{k} &  -\epsilon_{\mathbf{k}}  \\
 \end{pmatrix} . 
\end{align}
Let us assume for simplicity that $\phi_\text{a}$ and $\phi_\text{b}$ are real. Diagonalizing $h(\mathbf{k})$, the energy eigenvalues are found to be:
\begin{align}
E^1_\mathbf{k} & = \sqrt{  \epsilon_\mathbf{k}^2+\tilde{\epsilon}_\mathbf{k}^2+4\phi_\text{a}^2 +4\tilde{\epsilon}_\mathbf{k}\phi_\text{a}+4\phi_\text{b}^2 f_\mathbf{k}^2}\, ,\nn
E^2_\mathbf{k} & =  \sqrt{\epsilon_\mathbf{k}^2+\tilde{\epsilon}_\mathbf{k}^2+4\phi_\text{a}^2 -4\tilde{\epsilon}_\mathbf{k}\phi_\text{a}+ 4 \phi_\text{b}^2  f_\mathbf{k}^2} \,,\nn
E^3_\mathbf{k} & = -E^1_\mathbf{k}\, ,\quad
E^4_\mathbf{k} = - E^2_\mathbf{k} .
\end{align}
The ground state energy at zero temperature (same as free energy at $T=0$) becomes 
\begin{align}
F = \sum\limits_{n=1}^4 \int{[\rd\mathbf{k}]E_\mathbf{k}^n \theta(\mu-E_\mathbf{k}^n)} +\frac{4|\phi_\text{a}|^2}{g_\text{a}} + \frac{4|\phi_\text{b}|^2}{g_\text{b}} \, ,
\end{align}
where the sum over $n$ runs over four bands. For a given set of parameters, we only expect one ordered state. If one finds more than one ordered state, then the one with the lower free energy should dominate. Hence, for each ordered parameter, one can minimize the free energy to find a self-consistent equation. For example, for the SDW state, the self-consistent equation becomes 
\begin{align}
& \frac{8\phi_\text{a}}{g_\text{a}} = - \frac{\partial}{ \partial \phi_\text{a}}
\sum\limits_{n=1}^4  \int{[\rd\mathbf{k}] E_\mathbf{k}^n \theta(\mu-E_\mathbf{k}^n)}.
\end{align}
Similarly, one can derive self-consistency equations for the other phase as well by minimizing the free energy energy. We choose units such that $t=1$.

\begin{figure}[!b]
\centering
(a)\raisebox{-.7\height}{\includegraphics[width = 0.7\textwidth]{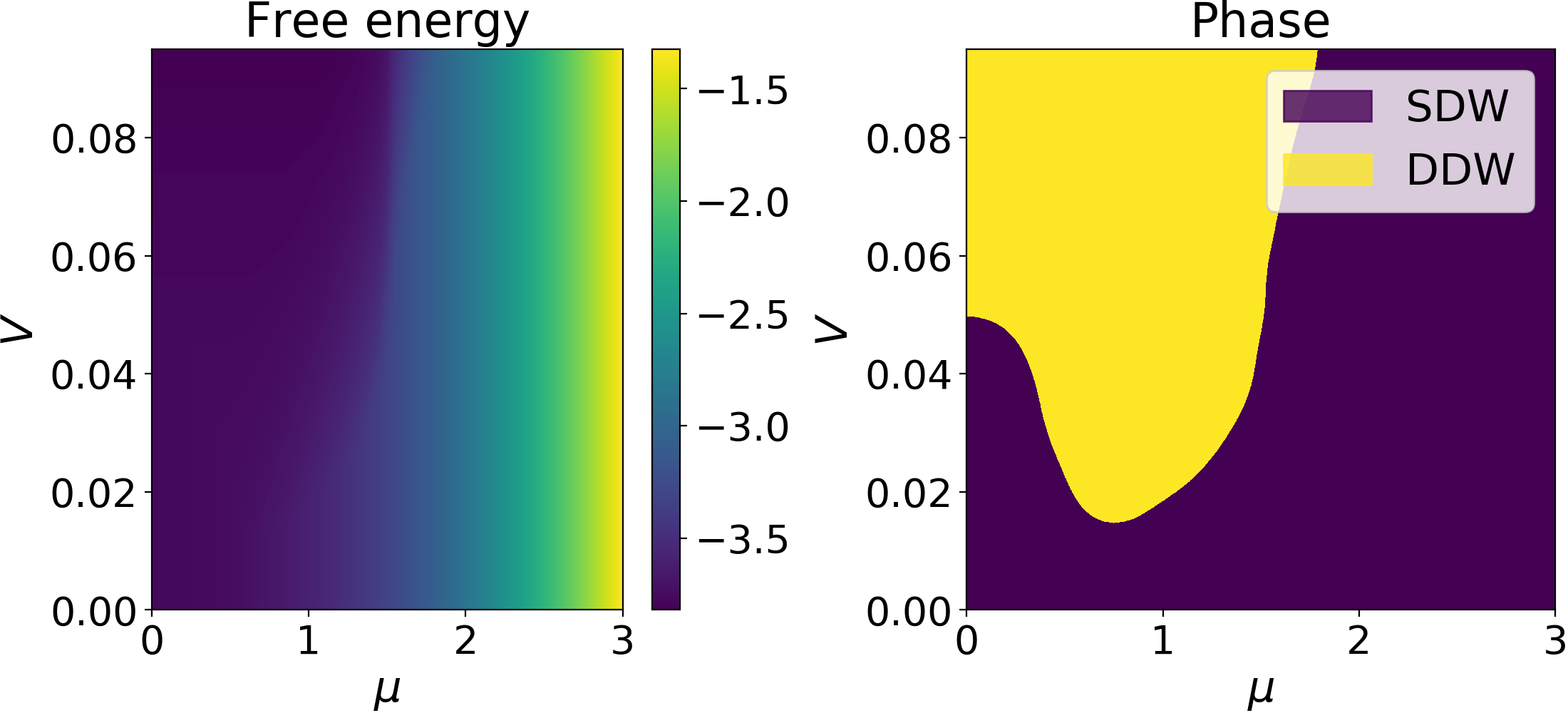}} \\
(b)\raisebox{-.7\height}{\includegraphics[width = 0.7\textwidth]{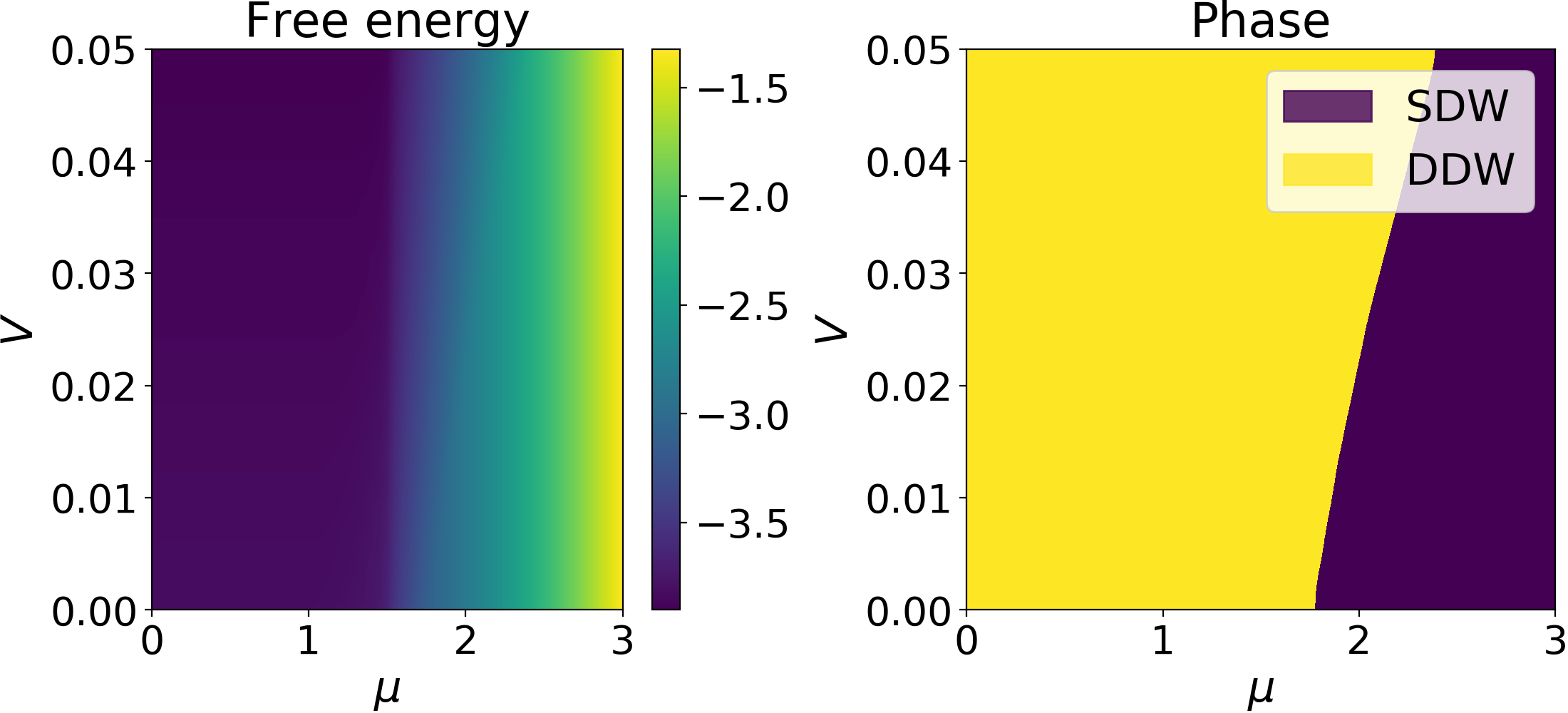}}
\caption{\label{fig1} (Colour online) The panels show the phase diagrams obtained for different values of $t_\text{c}$, in the $\mu{-}V$ plane. The QAH phase exists near $(\mu=0,  V=0)$ region in the last panel. We have set $t_\text{diag}=0.75$ and $U=1.0$. The values of $t_\text{c}$ are $ \{ 0.02,0.05\}$ for the successive panels in increasing order. The ranges for $V$ are such that the CWD phase can be neglected.
All the parameters are in units with $t=1$.}
\end{figure}

We note that the SDW vertex depends only on $U$, whereas the DDW vertex depends only on $V$ and $t_\text{c}$. Therefore, DDW phase will be favoured by increasing the values of $V$ and $t_\text{c}$. Clearly, if we want DDW phase to exist around $V=0$, we need to reach an optimum value of $t_\text{c}$. This is what our results show.

\begin{figure}[!t]
\centering
{\includegraphics[width =0.7\textwidth]{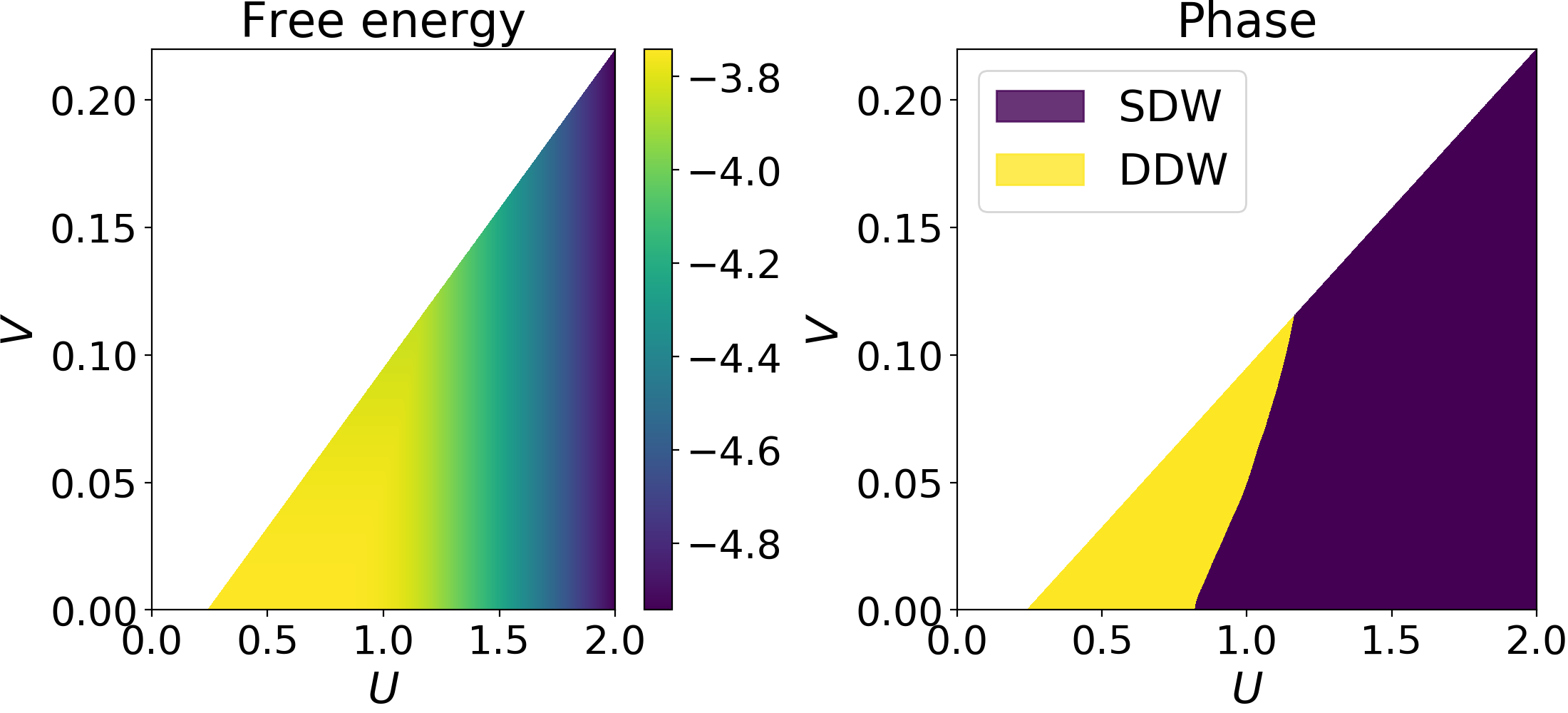}}
\caption{\label{fig2} (Colour online) Phase diagram for $\mu=0.1$, $t_\text{diag}=0.75$, $t_\text{c} =0.02$ in the $U{-}V$ plane for the allowed region in which the CDW phase can be neglected.}
\end{figure}

\begin{figure}[!t]
\vspace{2mm}
\centering
(a)\raisebox{-.7\height}
{\includegraphics[width =0.35  \linewidth]{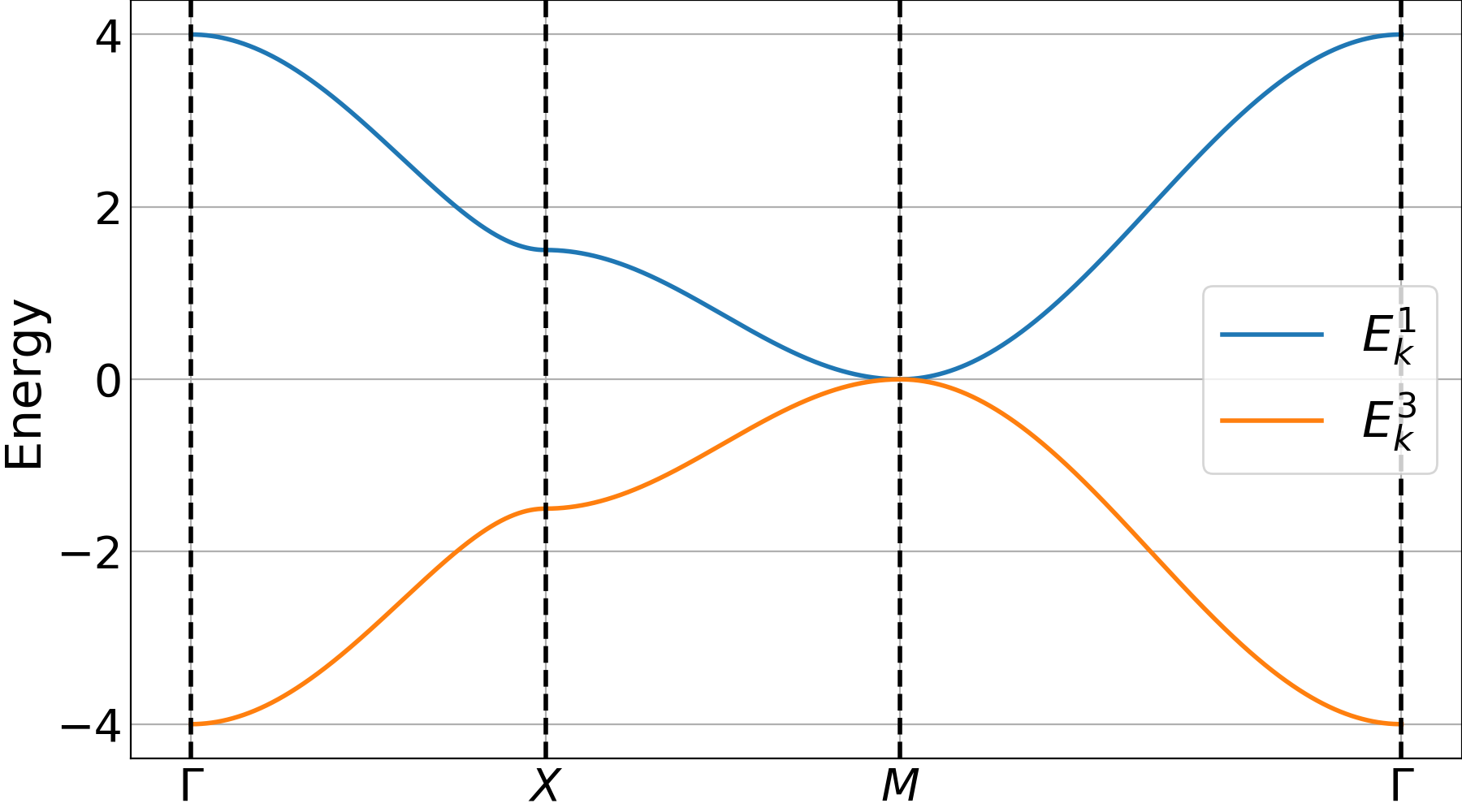}}
\hspace{1.5 cm}
(b)\raisebox{-.7\height}
{\includegraphics[width=0.35 \linewidth]{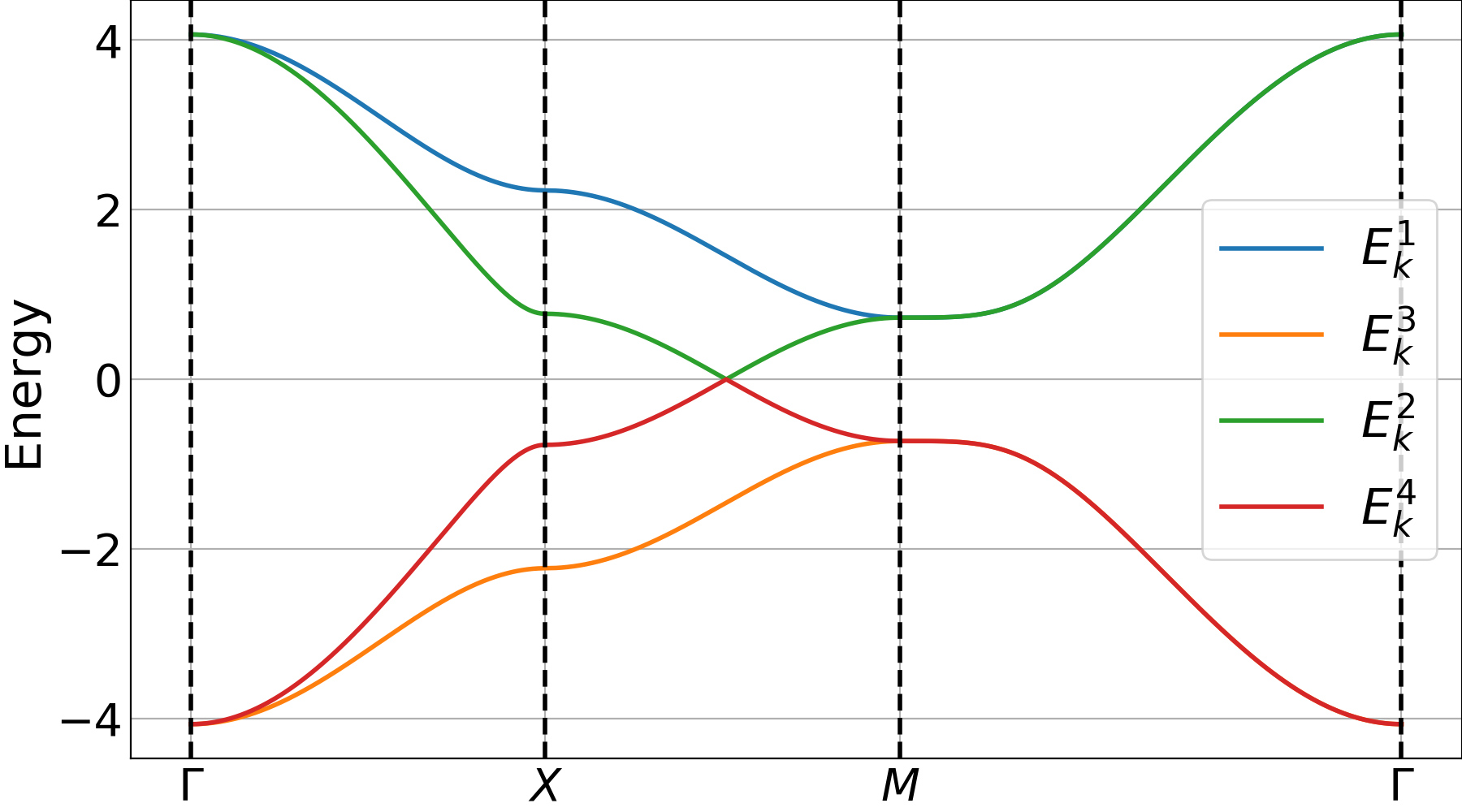}}\vspace{2mm}\\
(c)\raisebox{-.7\height}
{\includegraphics[width =0.35 \linewidth]{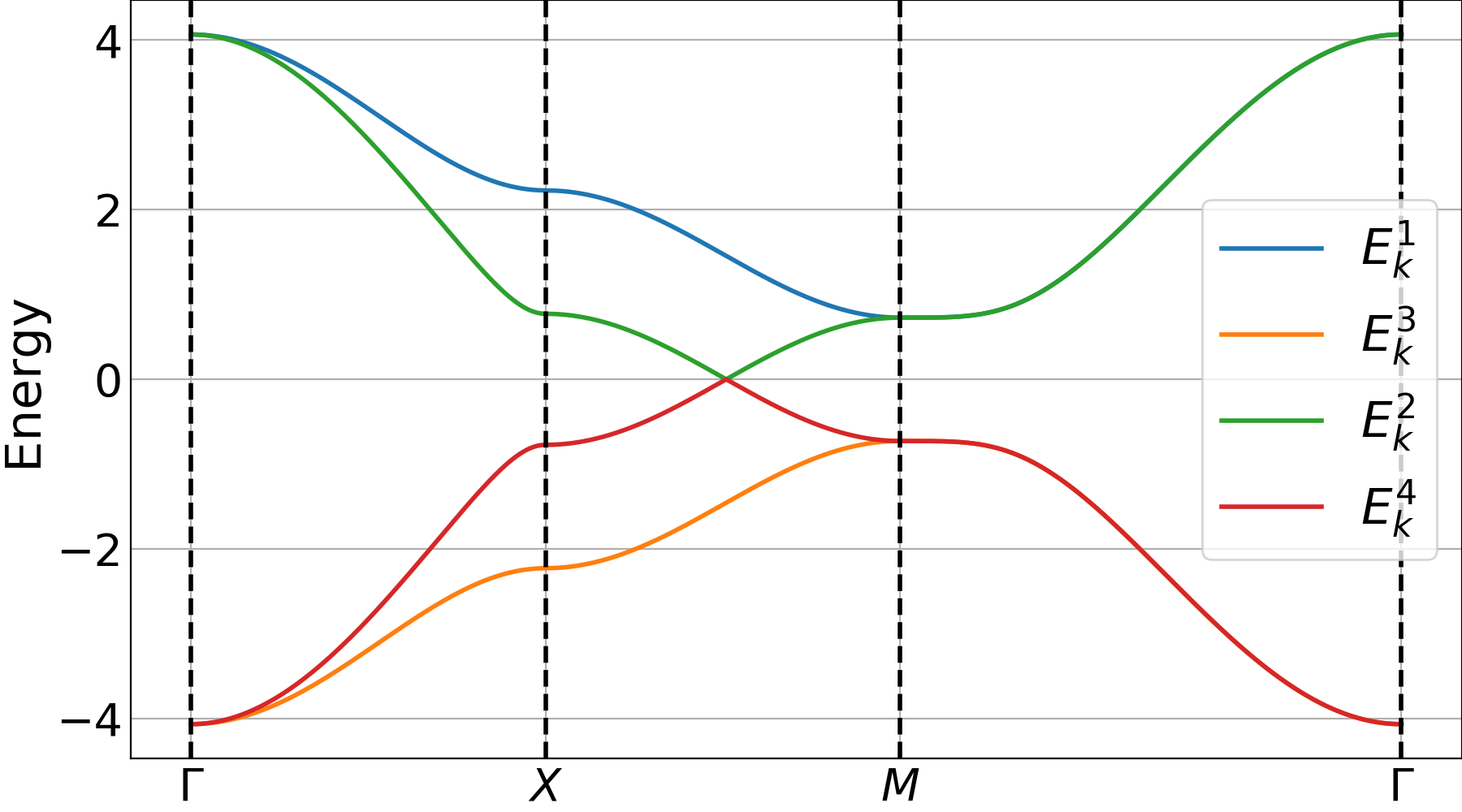}}
\hspace{1.5 cm}
(d)\raisebox{-.7\height}
{\includegraphics[width=0.35 \linewidth]{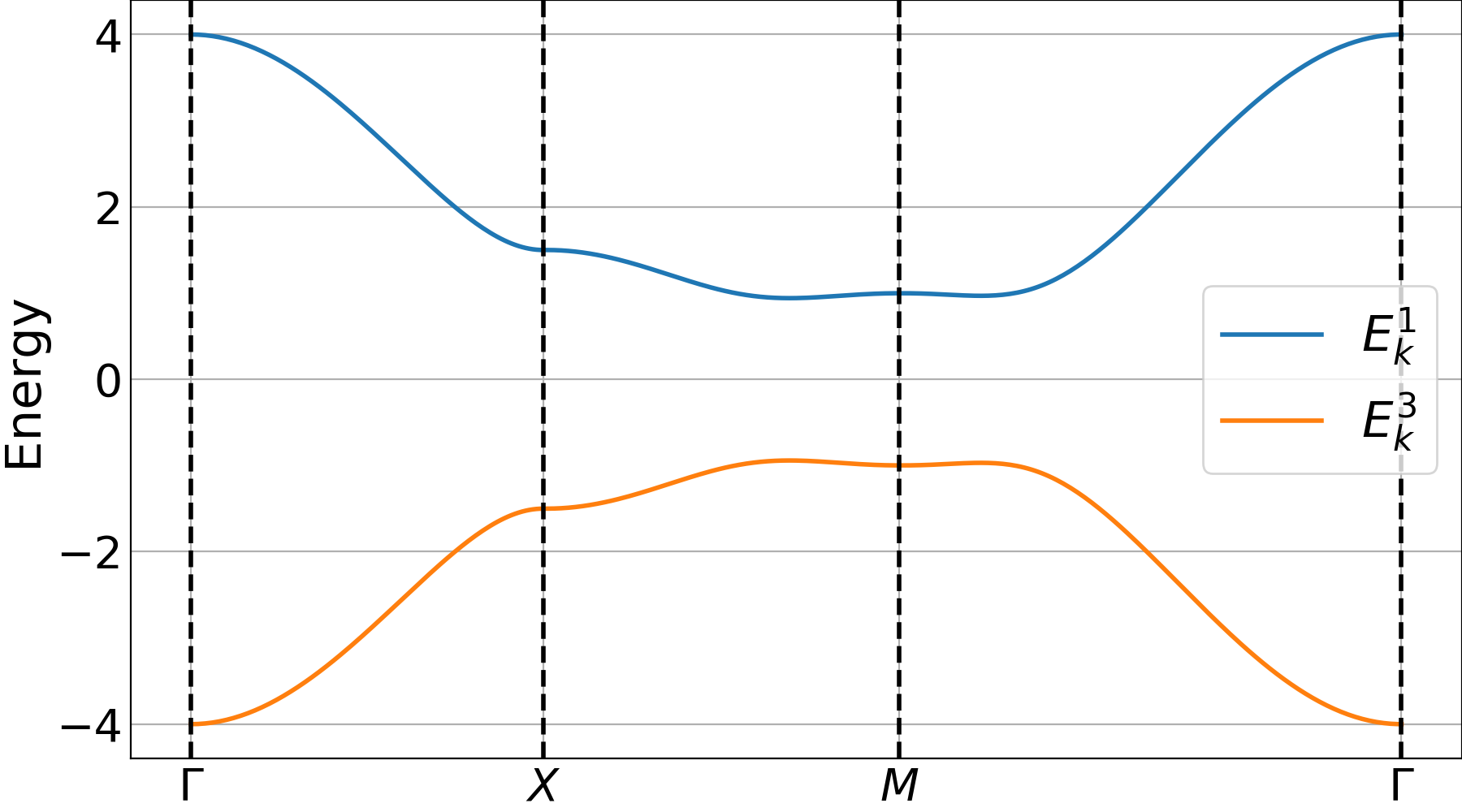}}\vspace{2mm}\\
(e)\raisebox{-.7\height}
{\includegraphics[width=0.35 \linewidth]{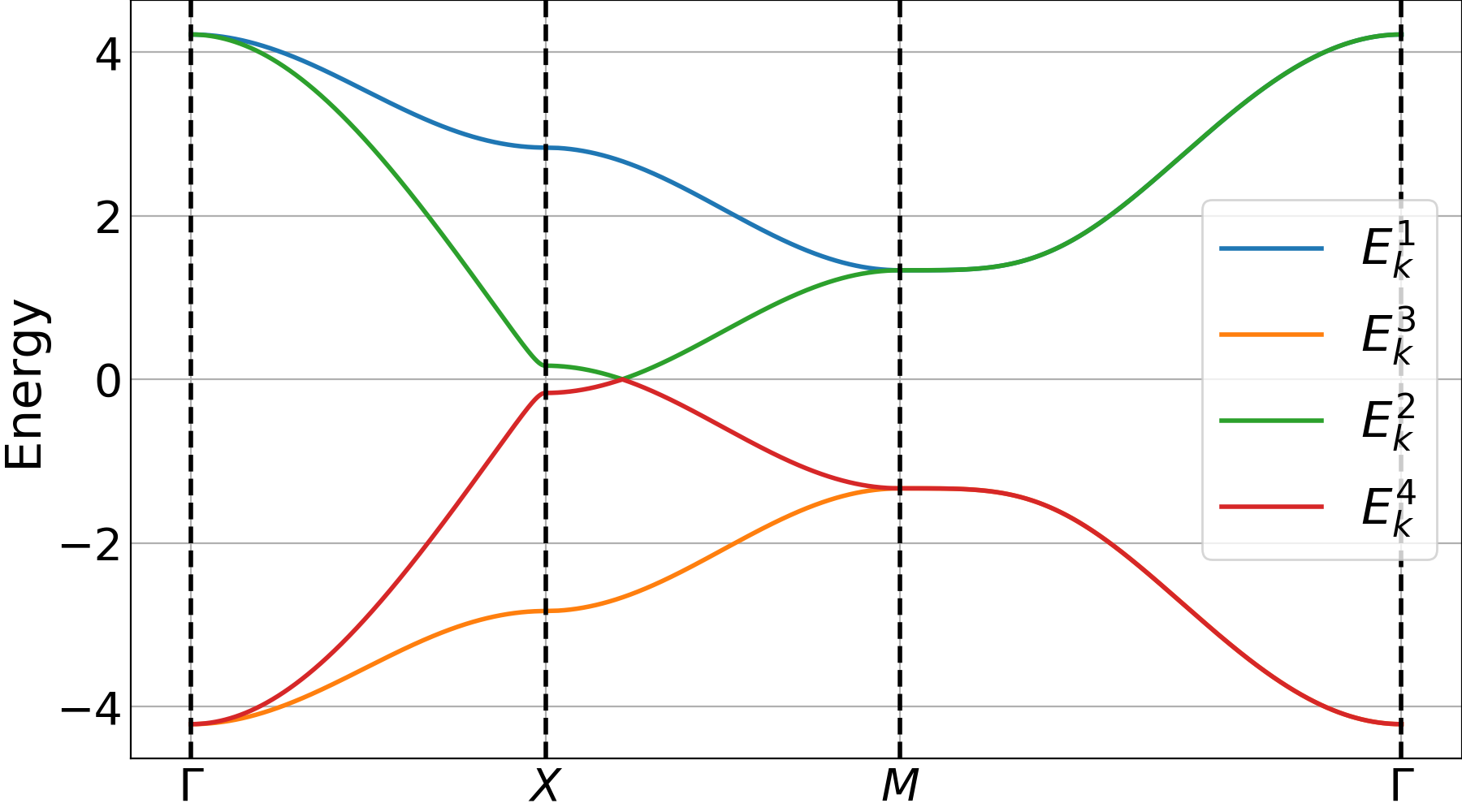}}
\hspace{1.5 cm}
(f)\raisebox{-.7\height}
{\includegraphics[width=0.35 \linewidth]{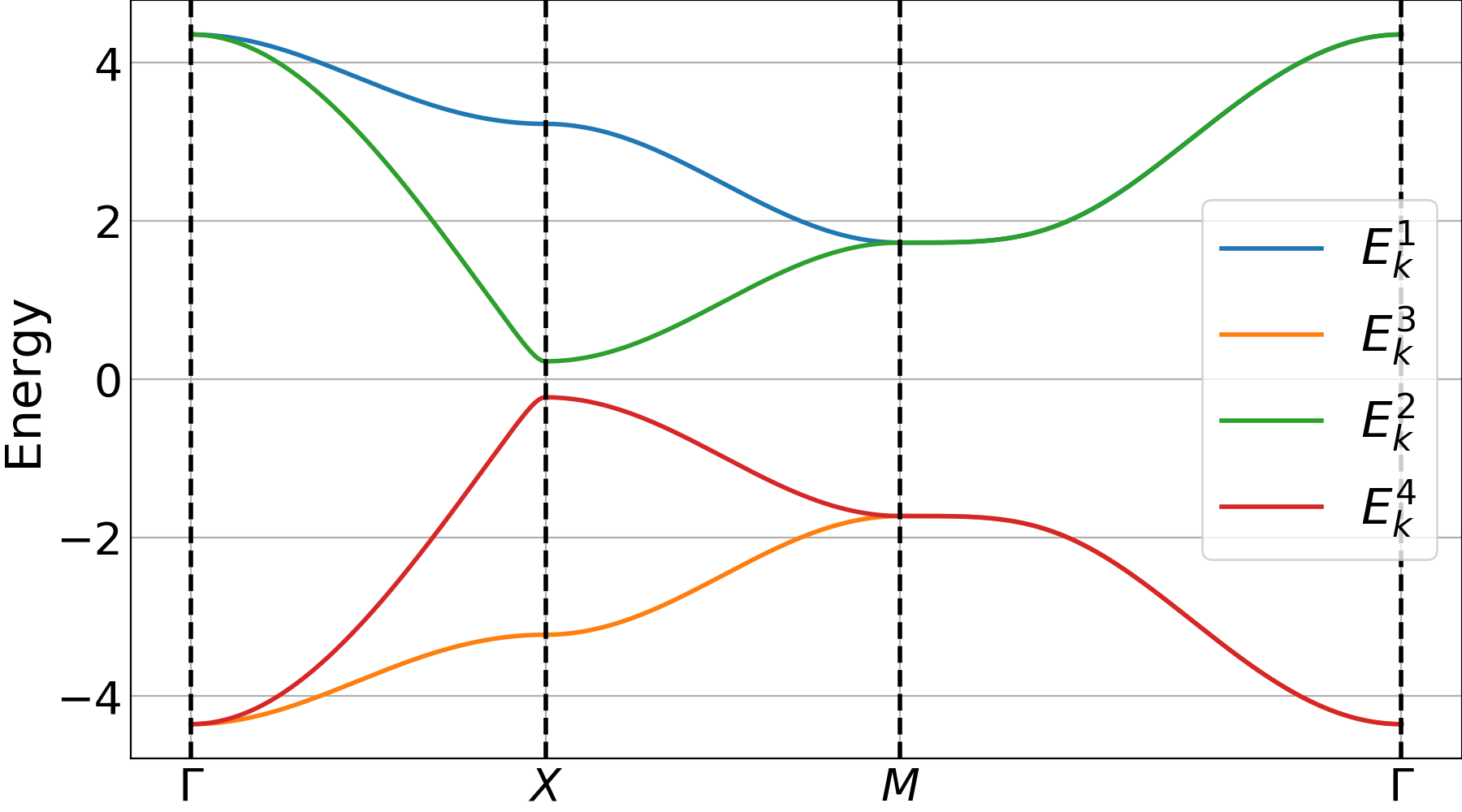}}\vspace{2mm}\\
(g)\raisebox{-.7\height}
{\includegraphics[width=0.35 \linewidth]{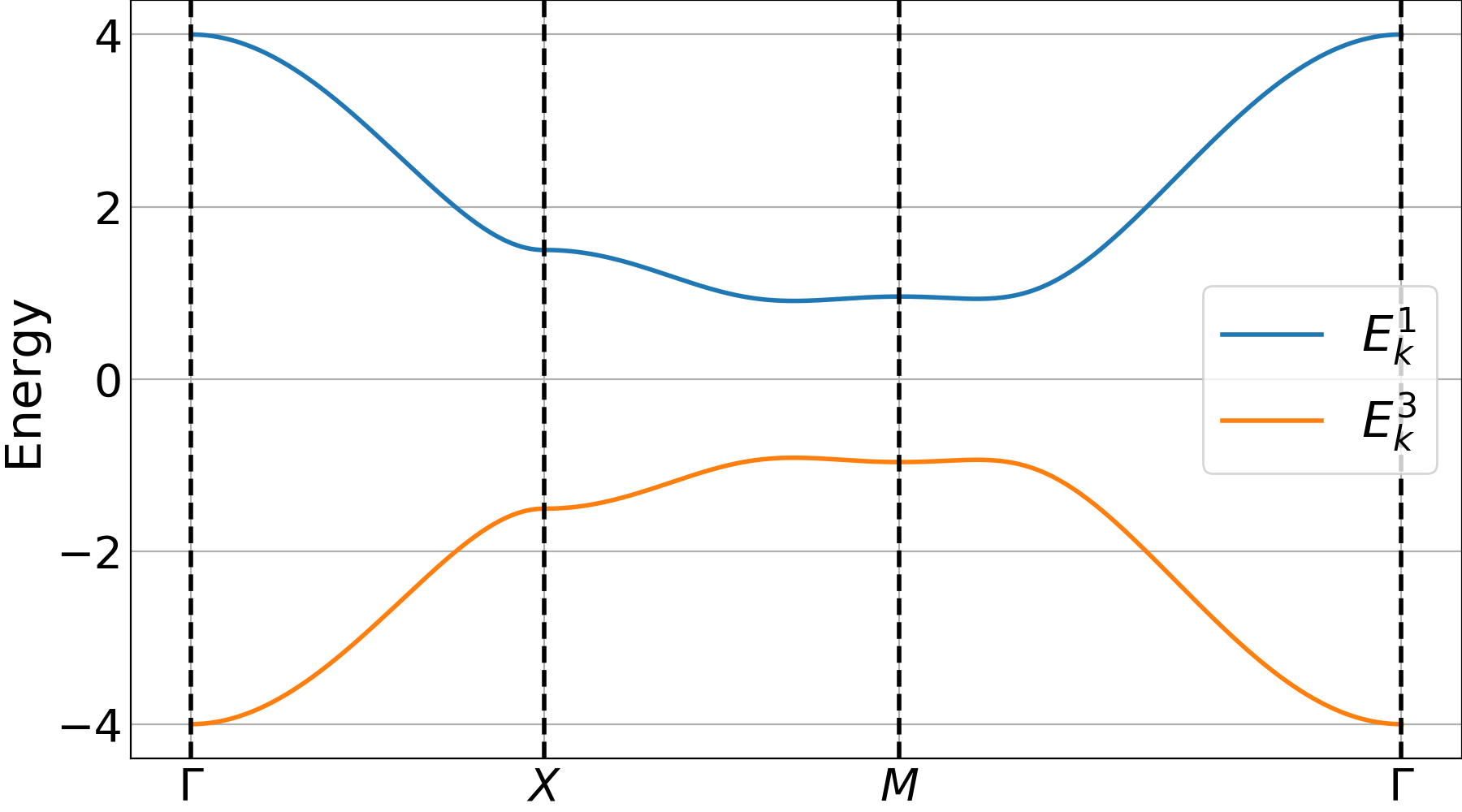}}
\hspace{1.5 cm}
(h)\raisebox{-.7\height}
{\includegraphics[width=0.35 \linewidth]
{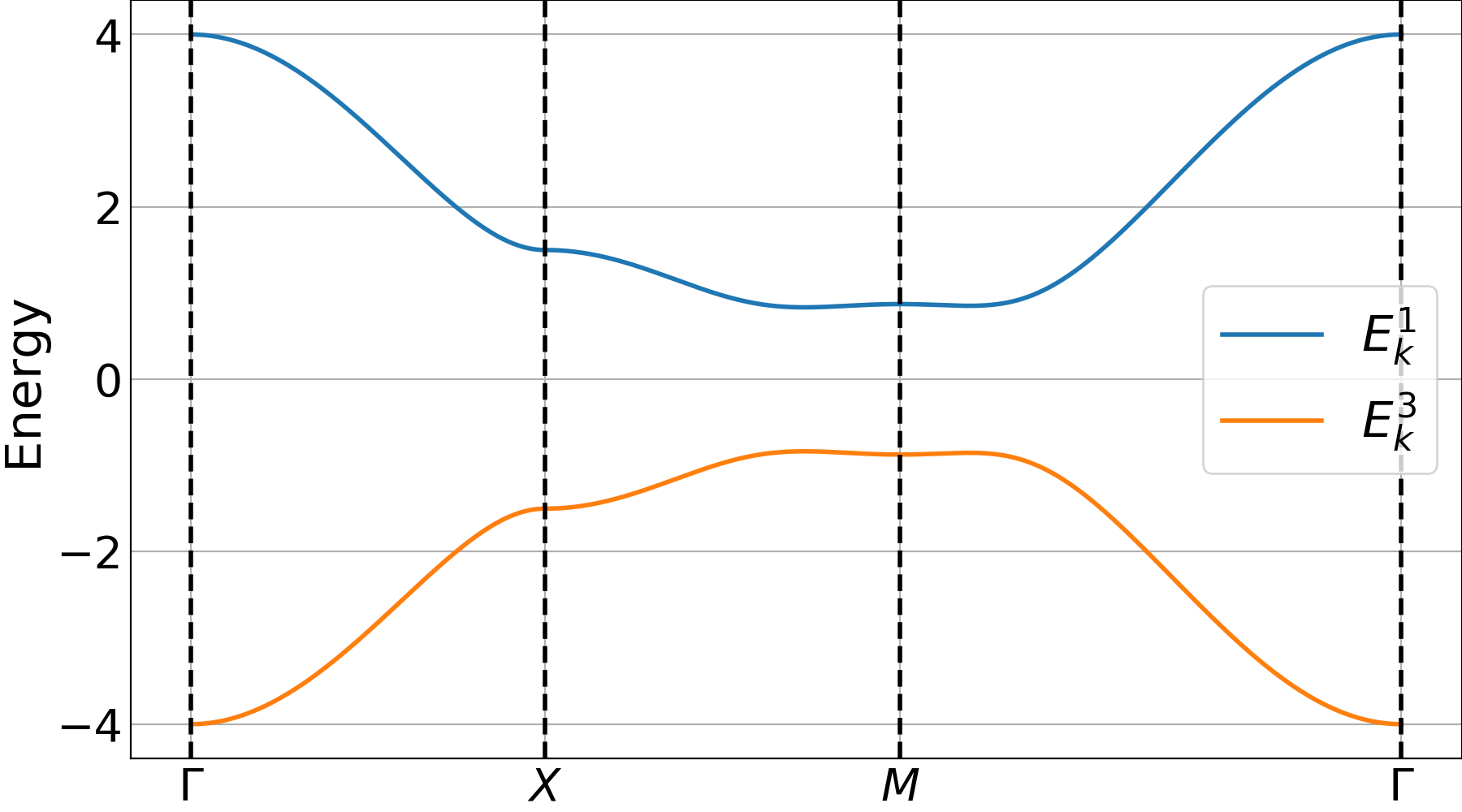}}
\caption{\label{fig3} (Colour online) The four energy bands for (a) $t_\text{diag}=0.75$,   $U=0$, $V=0$, $t_\text{c}=0$,  $\mu=0$ (non-interacting case); (b) $t_\text{diag}=0.75$,   $U=1$, $V=0$, $t_\text{c}=0.02$,  $\mu=0 $; (c) $t_\text{diag}=0.75$,  $U=1$, $V=0.04$, $t_\text{c}=0.02$,  $\mu=0 $; (d)~$t_\text{diag}=0.75$,  $U=14$, $V=0.02$, $t_\text{c}=0.05$,  $\mu=0 $; (e) $t_\text{diag}=0.75$,  $U=1$, $V=0.01$, $t_\text{c}=0.02$,  $\mu=1.0 $; (f) $t_\text{diag}=0.75$,  $U=1$, $V=0.01$, $t_\text{c}=0.02$,  $\mu=1.5 $; (g) $t_\text{diag}=0.75$,  $U=1$, $V=0.03$, $t_\text{c}=0.02$,  $\mu=1.0 $; (h) $t_\text{diag}=0.75$,  $U=1$, $V=0$, $t_\text{c}=0.05$,  $\mu=0 $. As usual, we use the convention: $\Gamma = (0,0)$,  $X = ( -\piup/2 , \piup/2)$, $M = ( 0, \piup )$, for the symmetry points of the RBZ. The panels showing only two energy bands actually involve each band being doubly degenerate.}
\end{figure}

\begin{figure}[!t]
\centering
(a)\raisebox{-.7\height}
{\includegraphics[width =0.4 \linewidth]{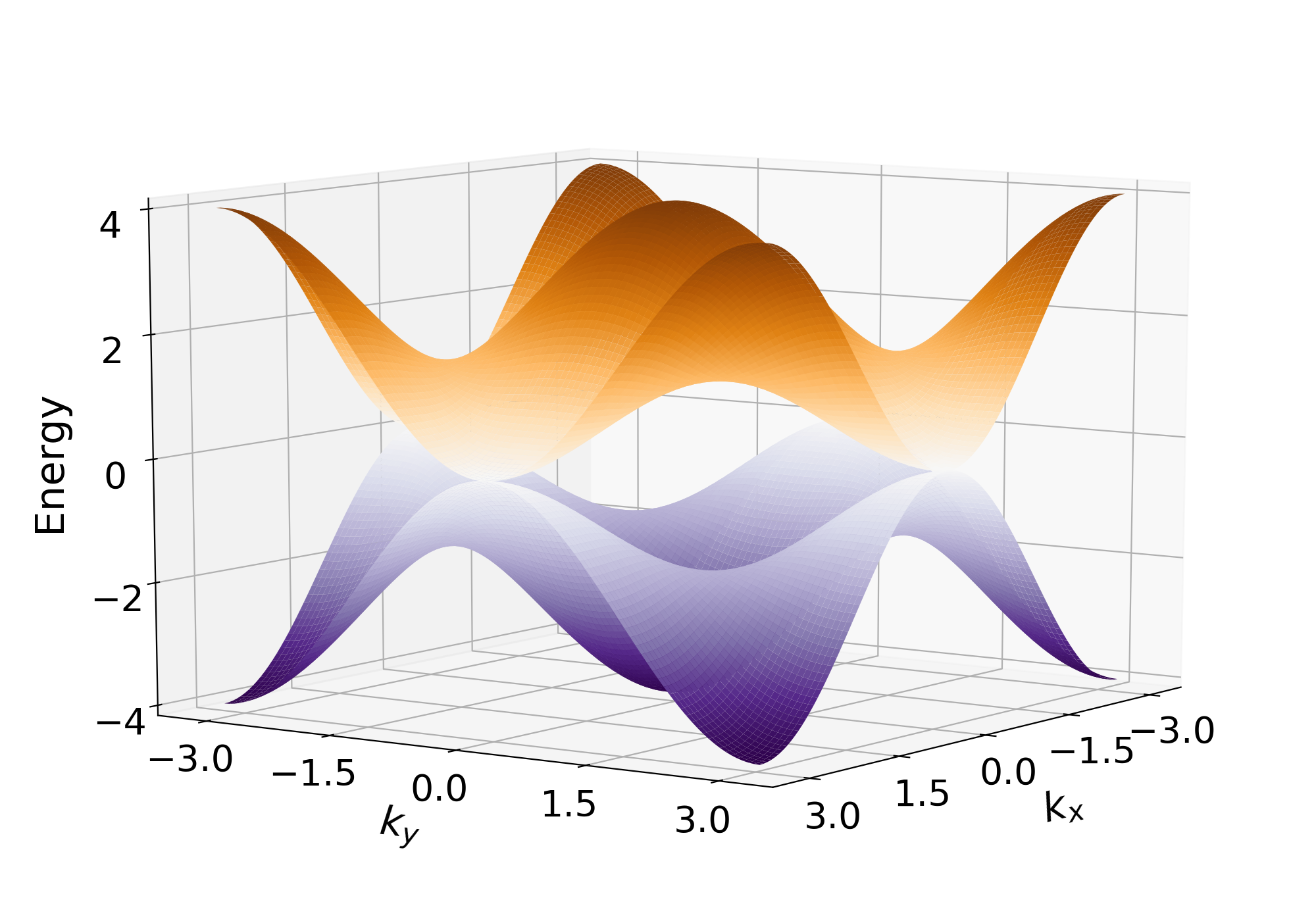}}
\quad
(b)\raisebox{-.7\height}
{\includegraphics[width=0.4 \linewidth]{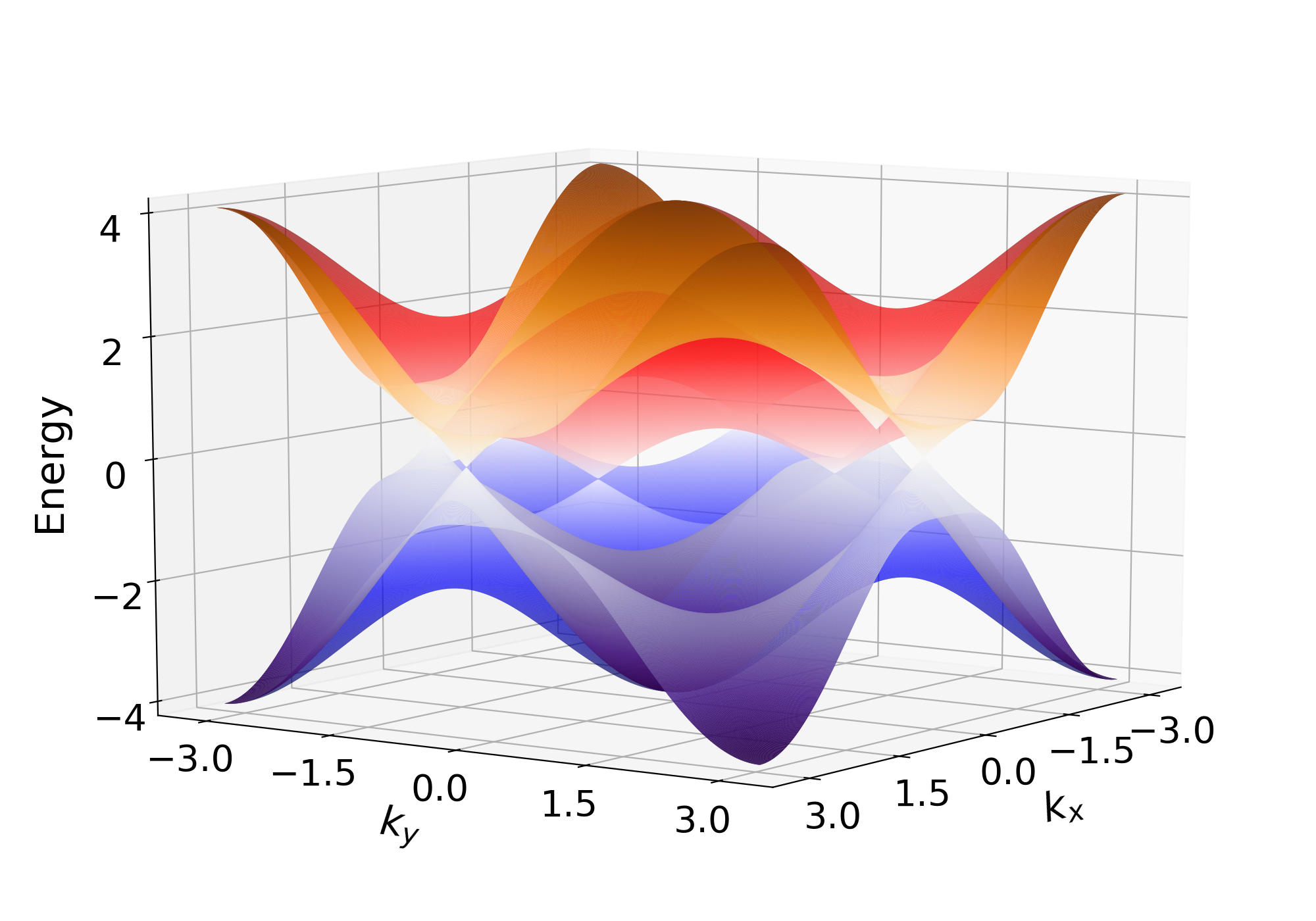}}
\quad
(c)\raisebox{-.7\height}
{\includegraphics[width=0.4 \linewidth]{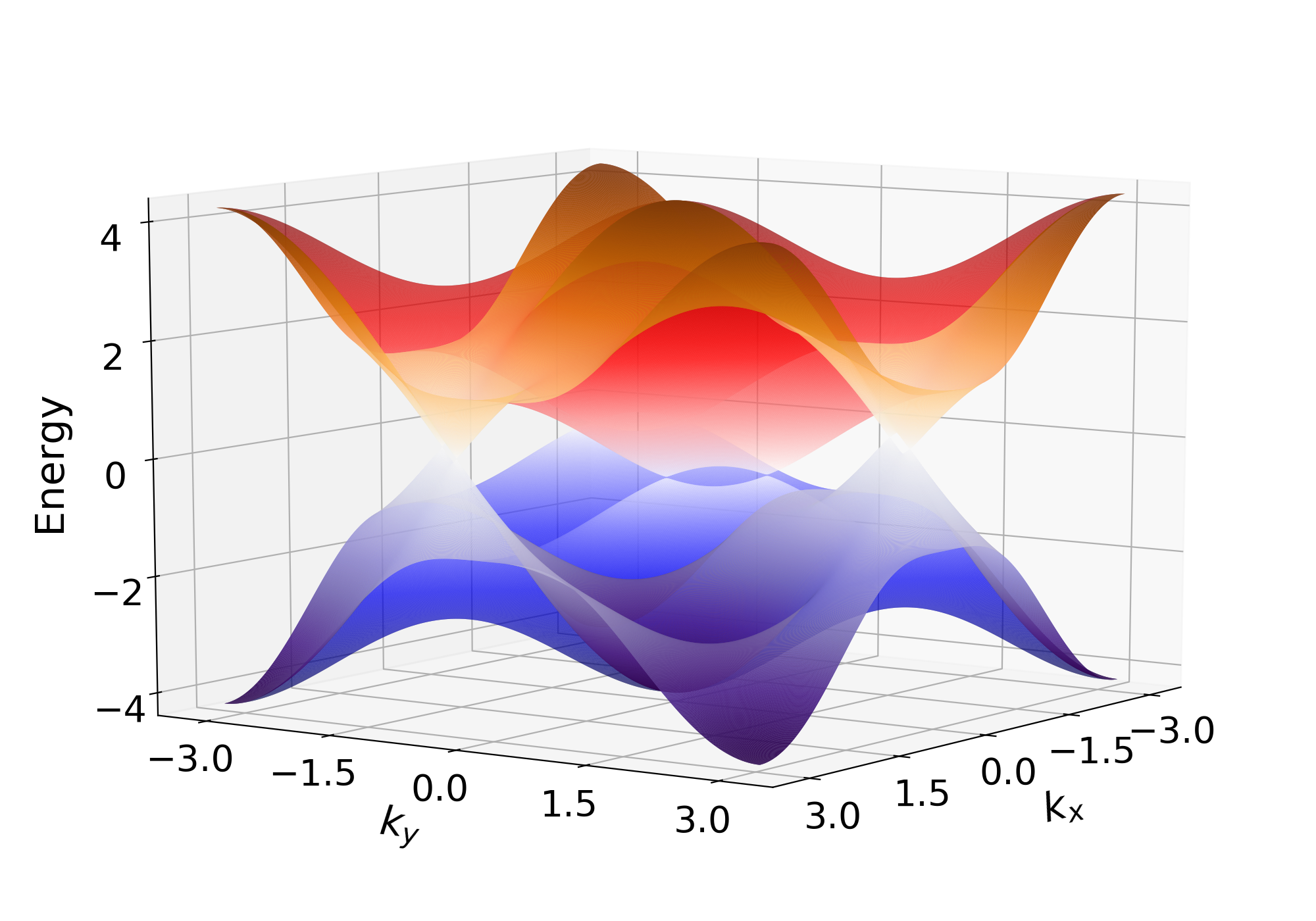}}
(d)\raisebox{-.7\height}
{\includegraphics[width=0.4 \linewidth]{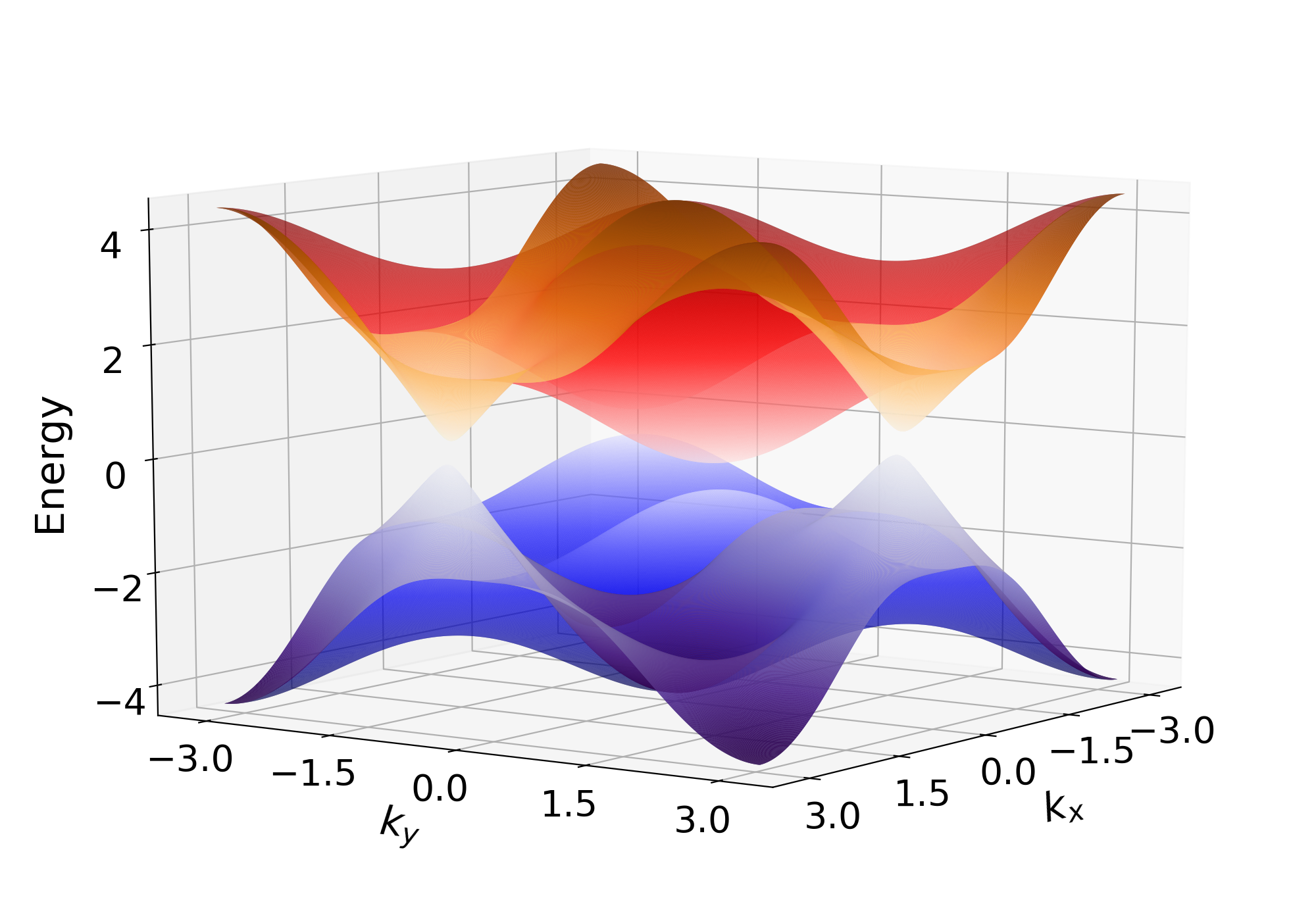}}
(e)\raisebox{-.7\height}
{\includegraphics[width=0.4 \linewidth]{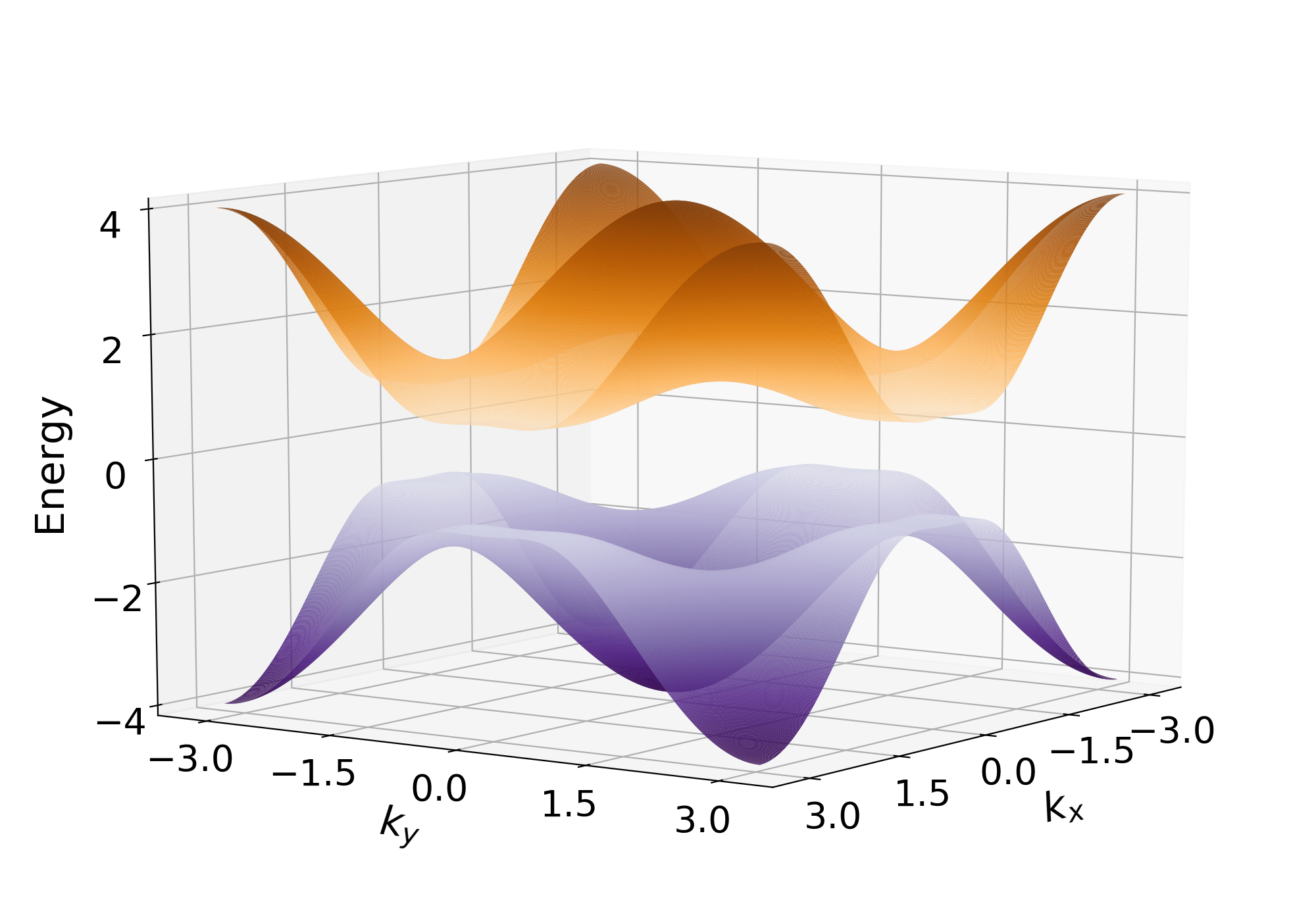}}
\quad
(f)\raisebox{-.7\height}
{\includegraphics[width=0.4 \linewidth]{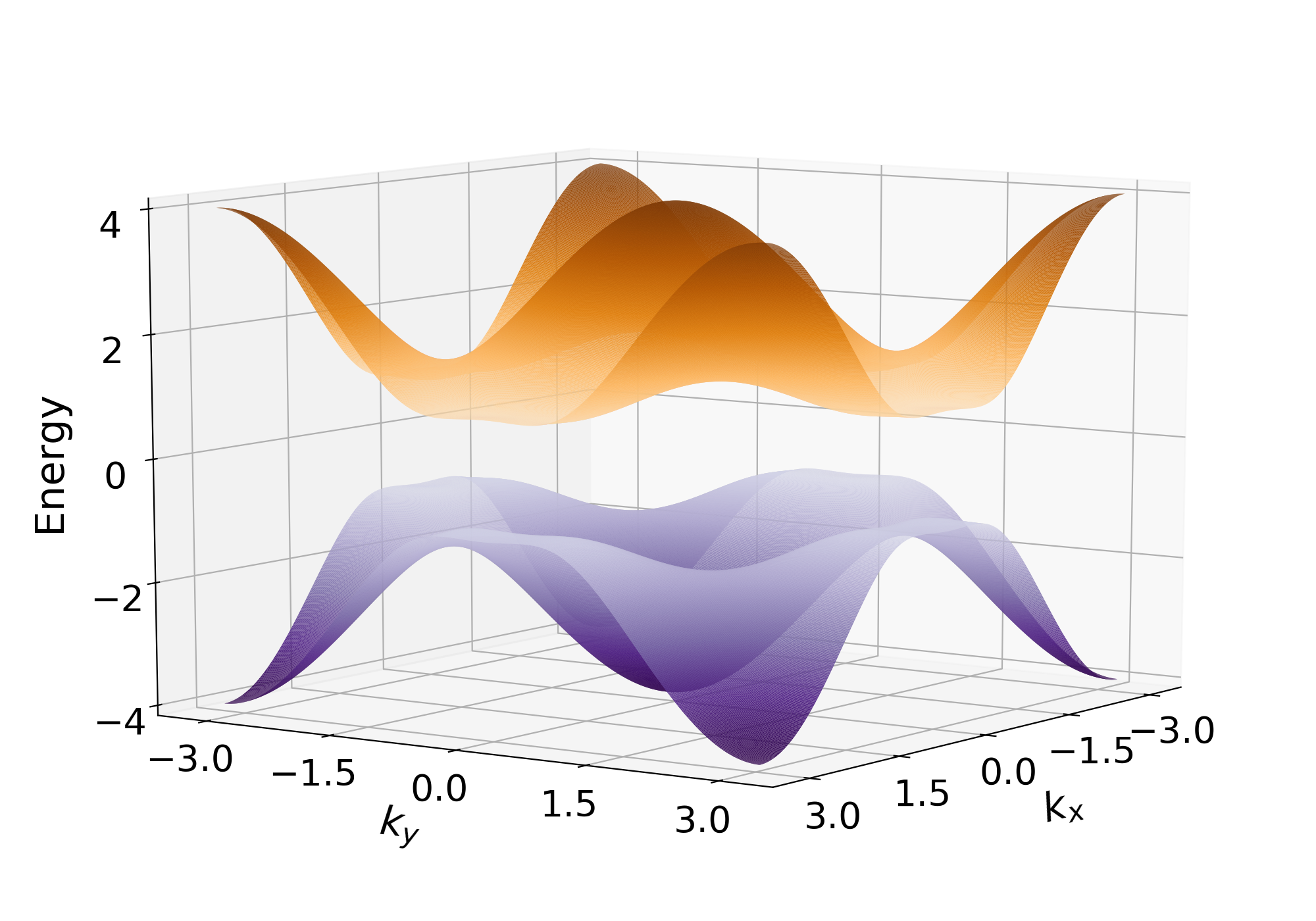}}
\caption{\label{fig4} (Colour online) The four energy bands for (a) $t_\text{diag}=0.75$,   $U=0$, $V=0$, $t_\text{c}=0$,  $\mu=0$ (non-interacting case); (b) $t_\text{diag}=0.75$,  $U=1$, $V=0.04$, $t_\text{c}=0.02$,  $\mu=0 $; (c) $t_\text{diag}=0.75$,  $U=1$, $V=0.01$, $t_\text{c}=0.02$,  $\mu=1.0 $; (d) $t_\text{diag}=0.75$,  $U=1$, $V=0.01$, $t_\text{c}=0.02$,  $\mu=1.5 $;  (e) $t_\text{diag}=0.75$,  $U=1$, $V=0$, $t_\text{c}=0.05$,  $\mu=0 $; (f) $t_\text{diag}=0.75$,  $U=1$, $V=0.03$, $t_\text{c}=0.02$,  $\mu=1.0 $. The panels showing only two energy bands actually involve each band being doubly degenerate.}
\end{figure}

Figure~\ref{fig1} shows the phase diagrams for $t_\text{diag}=0.75$ and $U =1.0$, such that the QAH phase appears around $(V=0,\mu=0)$ region by increasing $t_\text{c}$ to a nonzero optimum value, and extends continuously into the regions with $(V=0,\mu>0)$ and $(V>0,\mu>0)$. We have shown the ranges for $V$ for which we are allowed to neglect the CDW phase. 
Figure~\ref{fig2} is shown to emphasize that the QAH phase can indeed exist near $V=0 $ for a nonzero value of $\mu$. In this case, $t_\text{c}=0.05$ and only the allowed region for neglecting CDW vertex is shown in the phase diagram.
The energy bands are shown in figures~\ref{fig3} and \ref{fig4}, which indicate that interactions open up gaps at the quadratic band touching point. We note that for the QAH phase, the band opening is such that we have $E_{\mathbf k}^1 = E_{\mathbf k}^2$ and $E_{\mathbf k}^3 = E_{\mathbf k}^4$, where $E_{\mathbf k}^1 =- E_{\mathbf k}^3$. In other words, the QAH phase is characterised by two doubly degenerate bands which are negative of each other, similar to the non-interacting Hamiltonian energies [shown in figures~\ref{fig3}~(a) and \ref{fig4}~(a)]. This case is captured by the figures~\ref{fig3}~(d), \ref{fig3}~(g), \ref{fig3}~(h), \ref{fig4}~(e), \ref{fig4}~(f). On the contrary, for the SDW bands, there is no degeneracy. We note that when SDW appears at $\mu=0$ region, two of the bands still touch the other [see figures~\ref{fig3}~(b), \ref{fig3}~(c) and \ref{fig4}~(b)] --- a gap appears only at higher values of $\mu$ [see figures~\ref{fig3}~(e), \ref{fig3}~(f), \ref{fig4}~(c) and \ref{fig4}~(d)]. On the contrary, when QAH state appears at $\mu =0$, a gap appears between the positive and negative energy bands, each remaining doubly degenerate. Around $(V=0,\mu=0)$, a higher value of $t_\text{c}$ thus enables this gap-opening, making the QAH state feasible in that region. Since we have a large number of parameters in the theory, we have chosen to show the various possible scenarios with an exhaustive number of figures.

Our numerical results show that QAH/DDW phase can exist for nonzero $\mu$ (away from QBCP) in the interaction-driven scenario, both for $V =0$ and $V>0$. Furthermore, an optimum $t_\text{c}$ value allows QAH phase to exist around $(V=0,\mu=0)$. Hence, this extends the realm of current experimental investigations to find these topological phases.

Our simulations show that as $t_\text{c}$ and $V$ is increased, keeping the values of other coupling constants fixed, we move from an SDW phase around to a DDW phase $\mu =0$. That this will happen can also be seen from the relation $g_\text{DDW} = 8V+24 t_\text{c}$. This can be achieved in real experiments, for example, by applying pressure. Since correlated hopping appears from the off-diagonal elements of the Coulomb interactions between the nearest neighbouring lattice sites, pressure can affect the ease with which correlated hopping can take place by changing the lattice spacings. For example, there will be an increase in $t_\text{c}$ when the lattice spacing shrinks (under increased pressure) caused by increased admixtures of nearest neighbour electronic wavefunctions, whereas this will not affect the one-site Coulomb repulsion captured by $U$. Increased pressure will also tend to increase $V$.

%%%%%%%%%%%%%%%%%%%%%%%%%%%%%%%%%%%%%%%%%%%%%%%%%
\section{Conclusion}
\label{conclude}

We have shown by our numerically derived mean-field phase diagrams that the QAH (or DDW) state appears in a generic doping range without fine tuning in the presence of interactions of the right kind. Our interaction terms include correlated hopping, which essentially originates from strong local Coulomb repulsion. In future work, one can study the effect of disorder for the finite chemical potential case. 

Experimental investigations of systems showing QBCPs are just starting~\cite{Armitage}. Since it is experimentally challenging to tune exactly at the point of zero chemical potential, our work shows that the experimental explorations with extended realms can access topological Mott phases.

%%%%%%%%%%%%%%%%%%%%%%%%%%%%%%%%
\section*{Acknowledgements}
We acknowledge Sumanta Tewari for suggesting the problem. We thank Girish Sharma for collaboration in the initial stages of the project.
We also thank Matthias Punk, Nyayabanta Swain, Vito Scarola and Hoi Hui for fruitful discussions. Lastly, we express our deepest gratitude to Michael J. Lawler for carefully going through the manuscript and suggesting improvements.

\appendix

\section{DDW contribution from interactions}
\label{app}

In this appendix, we extract the DDW contribution from the interaction part, following \cite{foglio}. The spins are the same in this case and hence we drop the spin index. We replace the four-particle operators by sums of two-particle operators and C numbers:
\begin{align}
& c_1^\dagger c_2^\dagger c_3 c_4 \rightarrow \langle c_1^\dagger c_4\rangle  c_2^\dagger c_3
+ \langle c_2^\dagger c_3 \rangle  c_1^\dagger c_4  - \langle c_1^\dagger c_3 \rangle c_2^\dagger c_4
 - \langle c_2^\dagger c_4 \rangle   c_1^\dagger c_3 - \langle c_1^\dagger c_4\rangle \langle c_2^\dagger c_3 \rangle
 + \langle c_1^\dagger c_3 \rangle  \langle c_2^\dagger c_4 \rangle \nonumber
\\
\Rightarrow \ \
 &   c^\dagger _{\vec k_1}  c _{\vec k_2 } c^\dagger _{\vec k_ 3 }  c _{\vec k_4 } \rightarrow 
-\ \langle c_{\vec k_1}^\dagger c_{\vec k_4}\rangle c_{\vec k_3}^\dagger c_{\vec k_2}
- \langle c_{\vec k_3}^\dagger c_{\vec k_2} \rangle c_{\vec k_1}^\dagger c_{\vec k_4}  
+ \langle c_{\vec k_1}^\dagger c_{\vec k_2} \rangle c_{\vec k_3}^\dagger c_{\vec k_4 }
+  \langle c_{\vec k_3}^\dagger c_{\vec k_4 }  \rangle c_{\vec k_1}^\dagger c_{\vec k_2} \nn
&\hspace{ 2.28 cm} + \langle c_{\vec k_1}^\dagger c_{\vec k_4}\rangle  \langle c_{\vec k_3}^\dagger c_{\vec k_2} \rangle
 - \langle c_{\vec k_3}^\dagger c_{\vec k_4 }  \rangle  \langle  c_{\vec k_1}^\dagger c_{\vec k_2}\rangle .
\end{align} 
First, resolving into DDW mean-field ansatz, we get:
\begin{align}
& \delta ( \vec k_1+ \vec k_3- \vec k_2 -\vec k_4 )
c^\dagger _{\vec k_1}  c _{\vec k_2 } c^\dagger _{\vec k_ 3 }  c _{\vec k_4 }\nn \rightarrow \ \
&
-\ri \delta \left (  \vec k_3- \vec k_2 +\vec Q \right  ) \delta(\vec k_4 -\vec k_1+\vec Q)( \cos k_4^x- \cos k_4^y  )  c_{\vec k_3 }^\dagger  c_{\vec k_3  + \vec Q}
\nn 
&
- \ri   \delta \left (  \vec k_1- \vec k_4 +\vec Q \right  )  \delta(\vec k_2 -\vec k_3+\vec Q)( \cos k_2^x- \cos k_2^y  ) c_{\vec k_1 }^\dagger c_{\vec k_ 1 + \vec Q}   \nn
&
+ \ri \delta \left (  \vec k_3- \vec k_4 +\vec Q \right  ) \delta(\vec k_2 -\vec k_ 1+\vec Q) ( \cos k_2^x- \cos k_2^y  ) c_{\vec k_3 }^\dagger c_{\vec k_ 3 + \vec Q} 
\nn &
+ \ri \delta \left (  \vec k_1- \vec k_2 +\vec Q \right  ) \delta(\vec k_4 -\vec k_3+\vec Q)( \cos k_4^x- \cos k_4^y  ) c_{\vec k_1 }^\dagger c_{\vec k_1 + \vec Q} \nn
&
+ [\mbox{irrelevant terms involving }   ( \cos k_2^x- \cos k_2^y  ) ( \cos k_1^x- \cos k_1^y  ) \nn
&- ( \cos k_4^x- \cos k_4^y  )  ( \cos k_3^x- \cos k_3^y  ) ]
 %%%%%%%%%%%%%%%%%%%%%%%%%%
 \nn  \rightarrow \ \
 &
 -2\ri  \delta \left (  \vec k_3- \vec k_2 +\vec Q \right  ) \delta(\vec k_4 +\vec Q-\vec k_1)( \cos k_4^x- \cos k_4^y  )  c_{\vec k_3 }^\dagger  c_{\vec k_3  + \vec Q}
\nn &
[\mbox{interchanging dummy variables in second term as: } (\vec k_1, \vec k_2, \vec k_3, \vec k_4) \rightarrow  (\vec k_3, \vec k_4,\vec k_1,  \vec k_2)  ] \nn
%%%%%%%%%%%%%%%%%%%%%%%%%%%%%%%%%
&
+ 2\ri  \delta \left (  \vec k_3- \vec k_4 +\vec Q \right  ) \delta(\vec k_2 -\vec k_ 1+\vec Q) ( \cos k_2^x- \cos k_2^y  ) c_{\vec k_3 }^\dagger c_{\vec k_ 3 + \vec Q} 
\nn &
[\mbox{interchanging dummy variables in third term as: } (\vec k_1, \vec k_2, \vec k_3, \vec k_4) \rightarrow  (\vec k_3, \vec k_4,\vec k_1,  \vec k_2) ] .\label{coeff}
\end{align}
Multiplying by $ 2 V [\cos ( k^ x _3 - k^ x_4 )+\cos ( k^ y _3 - k^ y_4 ) ] $ and integrating over the appropriate variables, only the first term in the last two lines of equation~(\ref{coeff}) contributes, leading to the result:
\begin{align}
H_\text{int}  & = -2\ri \cdot 4 V \int  [\rd\vec k   ]
(\cos k_x - \cos k_y)   c^\dagger _{\vec k }  c _{\vec k +\vec Q }  \,,
\end{align}
where the factor $4$ is due to the choices $\pm \vec Q$.
Similarly, we can work out the correlated tunnelling part, for which the second term of equation~(\ref{coeff}) contributes. This gives
us the value of $g_\text{ddw} = 8  V +24 t_\text{c} $.

\section{Mean-field formulation for various order parameters}
\label{order}
%%%%%%%%%%%%%%%%%%%%%%%%%%%%%%%%%%%%%%

In this appendix, we demonstrate how we can formulate the mean-field theory for various order parameters.

\subsection{Triplet DDW}
First we consider the triplet $d_{x^2-y^2}$ ordered phase~\cite{Nayak:2000, Nayak:2002}, such that the Hamiltonian in equation~(\ref{ham-parent}) can be written as:
\begin{align}
H^\text{c}_\text{mf}= & -g_\text{c} \int [\rd\mathbf{k}][\rd\mathbf{k}']f_\mathbf{k} f_\mathbf{k'}
\left [ c^{\dagger}_{\mathbf{k}+\mathbf{Q}; \alpha}\sigma_z^{\alpha\beta}c_{\mathbf{k};\,\beta} \right ]
\left [ c^{\dagger}_{\mathbf{k}'; \gamma}  \sigma_z^{\gamma\delta}  c_{\mathbf{k}'+\mathbf{Q};\delta}  \right ],
\end{align}
where $g_\text{c}=g_\text{ddw}=8V+24t_\text{c}$. 
Expanding the spin indices above, we get:
%%%%%%%%%%%%55
\begin{align}
H^\text{c}_\text{mf} &=-g_\text{c}\int  [\rd\mathbf{k}]   [\rd\mathbf{k}']f_\mathbf{k} f_\mathbf{k}'\left  [ c^{\dagger}_{\mathbf{k}+\mathbf{Q} ;\uparrow }  c_{\mathbf{k};\uparrow } - c^{\dagger}_{\mathbf{k} +\mathbf{Q};\downarrow} c_{\mathbf{k};\downarrow}  \right ] \left  [ c^{\dagger}_{\mathbf{k'} ;\uparrow }  c_{\mathbf{k'}+\mathbf{Q};\uparrow } - c^{\dagger}_{\mathbf{k'} ;\downarrow} c_{\mathbf{k'}+\mathbf{Q};\downarrow}  \right ]\nn
%%%%%%%%%%%
&=-g_\text{c}\int  [\rd\mathbf{k}][\rd\mathbf{k}']f_\mathbf{k} f_\mathbf{k}'
\Big [ c^{\dagger}_{\mathbf{k}+\mathbf{Q} ;\uparrow }  c_{\mathbf{k};\uparrow }c^{\dagger}_{\mathbf{k'} ;\uparrow }  c_{\mathbf{k'} +\mathbf{Q};\uparrow }
 -  c^{\dagger}_{\mathbf{k}+\mathbf{Q} ;\uparrow }  c_{\mathbf{k};\uparrow }c^{\dagger}_{\mathbf{k'} ;\downarrow }  c_{\mathbf{k'} +\mathbf{Q};\downarrow }
 %%%%%%%%%%%%%%%%%%%%%%%%%%%%%%%%%
\nn & \quad
-c^{\dagger}_{\mathbf{k}+\mathbf{Q} ;\downarrow }  c_{\mathbf{k};\downarrow }c^{\dagger}_{\mathbf{k'} ;\uparrow }  c_{\mathbf{k'} +\mathbf{Q};\uparrow }
 +  c^{\dagger}_{\mathbf{k}+\mathbf{Q} ;\downarrow }  c_{\mathbf{k};\downarrow }c^{\dagger}_{\mathbf{k'} ;\downarrow }  c_{\mathbf{k'} +\mathbf{Q};\downarrow }
\Big ] .
\end{align}
We define the two mean-field order parameters:
\begin{align}
&\phi_\text{c}^ {\uparrow} =  g_\text{c}\int{[\rd\mathbf{k}]f_{\mathbf{k}}\langle c^\dagger_{\mathbf{k}+\mathbf{Q};\uparrow}}c_{\mathbf{k};\uparrow}\rangle, \quad
\phi_\text{c}^{\downarrow} = g_\text{c}\int{[\rd\mathbf{k}]f_{\mathbf{k}}\langle c^\dagger_{\mathbf{k}+\mathbf{Q};\downarrow}} c_{\mathbf{k};\downarrow}\rangle,
\end{align}
and expand the four-fermion operators using these to obtain the mean-field Hamiltonian: 
\begin{align}
H_\text{mf}^\text{ c} &=  -\int [\rd\mathbf{k}] \bigg [ \phi_\text{c}^\uparrow f_\mathbf{k}c_{\mathbf k; \uparrow}^\dagger
 c_{\mathbf k+\mathbf{Q}; \uparrow}+\phi_\text{c}^{\uparrow*}f_\mathbf{k}c_{\mathbf k+\mathbf{Q}; \uparrow}^\dagger c_{\mathbf k; \uparrow}  - \frac{\phi_\text{c}^\uparrow \phi_\text{c}^{\uparrow*}}{g_\text{c}}
%%%%%%%%%%%%%
 -\phi_\text{c}^\uparrow f_\mathbf{k} c_{\mathbf k; \downarrow}^\dagger c_{\mathbf k+\mathbf{Q}; \downarrow}-\phi_\text{c}^{\downarrow*}f_\mathbf{k}c_{\mathbf k+\mathbf{Q}; \uparrow}^\dagger c_{\mathbf k; \uparrow}  \nn
%%%%%%%%%%%%%%%%%%%%%%%%%%%%%%
& + \frac{\phi_\text{c}^\uparrow  \phi_\text{c}^{\downarrow*}}{g_\text{c}}-\phi_\text{c}^\downarrow f_\mathbf{k} c_{\mathbf k; \uparrow}^\dagger c_{\mathbf k+\mathbf{Q}; \uparrow}-\phi_\text{c}^{\uparrow*}f_\mathbf{k} c_{\mathbf k+\mathbf{Q}; \downarrow}^\dagger c_{\mathbf k; \downarrow}  + \frac{\phi_\text{c}^{\uparrow*}\phi_\text{c}^{\downarrow}}{g_\text{c}}
%%%%%%%%%%%%%%%%%%%%
 + \phi_\text{c}^\downarrow f_\mathbf{k}c_{\mathbf k; \downarrow}^\dagger c_{\mathbf k+\mathbf{Q}; \downarrow}+\phi_\text{c}^{\downarrow*}f_\mathbf{k} c_{\mathbf k+\mathbf{Q}; \downarrow}^\dagger c_{\mathbf k; \downarrow} \nn 
 &- \frac{\phi_\text{c}^\downarrow \phi_\text{c}^{\downarrow*}}{g_\text{c}}  \bigg ].
\end{align}
We will now choose $\phi_\text{c}^\uparrow=\ri\phi_\text{c}$, and $\phi_\text{c} ^\downarrow=-\ri\phi_\text{c}$, where $\phi_\text{c} \in \mathbb{R} $, as this choice gives spin currents of equal magnitude (QAH state). The mean-field Hamiltonian then takes the form:
\begin{align}
H_\text{mf}^ \text{c} &=\int [\rd{\mathbf{k}}] \psi^\dagger_{\mathbf{k}}  \begin{pmatrix}
  \epsilon_{\mathbf{k}}-\mu &\tilde{\epsilon}_\mathbf{k}+2\ri \phi_\text{c} f_\mathbf{k} & 0 & 0\\
  \tilde{\epsilon}_\mathbf{k}-2\ri \phi_\text{c} f_\mathbf{k} & -\epsilon_{\mathbf{k}}-\mu & 0 & 0\\
  0 & 0 &\epsilon_{\mathbf{k}}-\mu & \tilde{\epsilon}_\mathbf{k}-2\ri\phi_\text{c}f_\mathbf{k}\\
  0 & 0 & \tilde{\epsilon}_\mathbf{k}+2\ri\phi_\text{c}f_\mathbf{k} &  -\epsilon_{\mathbf{k}} -\mu\\
 \end{pmatrix}    \psi_{\mathbf{k}}
 + \frac{4\phi_\text{c}^2}{g_\text{c}}\,,\nn
\psi^\dagger_{\mathbf{k}}  & =\left (c^\dagger_{\mathbf{k}; \uparrow}, c^\dagger_{\mathbf{k}+\mathbf{Q}; \uparrow}, c^\dagger_{\mathbf{k}; \downarrow}, c^\dagger_{\mathbf{k}+\mathbf{Q}; \downarrow}  \right ) .
\end{align}
%%%%%%%%%%%%%%%%55

\subsection{SDW and singlet DDW}

Using the expressions in equation~(\ref{defn}) for the SDW and the singlet $d_{x^2-y^2}$ order parameters,
where the corresponding couplings are $g_\text{a}=g_\text{sdw}=2U$ and $g_\text{b}=g_\text{ddw}=8V+24t_\text{c}  $,
one can find the mean-field Hamiltonian for these order parameters in the manner outlined for triplet DDW.

\ukrainianpart

\title{Поява топологічних мотівських діелектриків поблизу точок дотику квадратичної зони}
\author{І. Мандал\refaddr{label1}, С. Гемшейм\refaddr{label2}}
\addresses{
\addr{label1} Лабораторія фізики атома і твердого тіла, Корнельський університет, Ітака, NY 14853, США 
\addr{label2} Фізичний факультет, технічний університет Дрездена, 01069 м. Дрезден, Німеччина
}

\makeukrtitle

\begin{abstract}
Нещодавно, у галузі фізики сильно скорельованих електронів розпочались інтенсивні дослідження кореляційно індукованої топологічної діелектричної фази.
Прикладом може слугувати точка дотику квадратичної зони, яка виникає на ґратці типу шахівниці при напівзаповненні, і в присутності взаємодій
призводить до появи  топологічних мотівських діелектриків.
В цій роботі, ми здійснили обчислення в наближенні середнього поля, щоб показати, що дана система демонструє нестійкість  по відношенню до топологічних
діелектричних фаз навіть далеко від напівзаповнення  (хімічний потенціал $\mu  =  0 $). Параметри взаємодії включають в себе одновузлове відштовхування ($ U $), 
відштовхування між найближчими сусідами ($ V $) та  наступні за найближчими сусідами скорельовані стрибки ($ t_\text{c} $). 
Взаємодія  $t_\text{c}$ виникає завдяки сильному кулонівському відштовхуванню. 
Регулюючи значення цих параметрів, ми отримали бажану топологічну фазу, яка охоплює область в межах  $(V  =  0  ,  \mu  =  0)$, розповсюджуючись
до областей з  $(V>0,\mu=0)$ та $(V>0,\mu>0)$. Це дає змогу розширити область поточних експериментальних зусиль з метою знаходження цих топологічних фаз.
\keywords шахові ґрати, точки дотику квадратичної зони, топологія, мотівський діелектрик, $d$-густинна хвиля
\end{abstract}

\end{document}